\documentclass[prd,aps,showpacs,nofootinbib,preprint,eqsecnum]{revtex4}
\usepackage{graphicx,color,amsmath,amsxtra}
\usepackage{amssymb}
\usepackage[english]{babel}
\usepackage{amsfonts}
\allowdisplaybreaks[4]

\newcommand{\beq}{\begin{equation}}
\newcommand{\eeq}{\end{equation}}
\newcommand{\bea}{\begin{eqnarray}}
\newcommand{\eea}{\end{eqnarray}}
\newcommand{\bseq}{\begin{subequations}}
	\newcommand{\eseq}{\end{subequations}}

\newcommand{\Ref}[1]{(\ref{#1})}

\begin{document}

\title{ General dynamical properties of cosmological models with nonminimal kinetic coupling}

\author{Jiro Matsumoto\footnote{E-mail address: jmatsumoto@kpfu.ru}
and Sergey V.~Sushkov\footnote{E-mail address: sergey\_sushkov@mail.ru}}
\affiliation{Institute of Physics, Kazan Federal University, Kremlevskaya Street 16a,
Kazan 420008, Russia}
	
\begin{abstract}
We consider cosmological dynamics in the theory of gravity with the scalar field possessing the nonminimal kinetic coupling to curvature given as $\eta G^{\mu\nu}\phi_{,\mu}\phi_{,\nu}$, where $\eta$ is an arbitrary coupling parameter, and the scalar potential $V(\phi)$ which assumed to be as general as possible.
With an appropriate dimensionless parametrization we represent the field equations as an autonomous dynamical system which contains ultimately only one arbitrary function $\chi (x)= 8 \pi \vert \eta \vert V(x/\sqrt{8 \pi})$ with $x=\sqrt{8 \pi}\phi$.
Then, assuming the rather general properties of $\chi(x)$, we analyze stationary points and their stability, as well as all possible asymptotical regimes of the dynamical system.
It has been shown that for a broad class of $\chi(x)$ there exist attractors representing three accelerated regimes of the Universe evolution, including de Sitter expansion (or late-time inflation), the Little Rip scenario, and the Big Rip scenario.
As the specific examples, we consider a power-law potential $V(\phi)=M^4(\phi/\phi_0)^\sigma$, Higgs-like potential $V(\phi)=\frac{\lambda}{4}(\phi^2-\phi_0^2)^2$, and exponential potential $V(\phi)=M^4 e^{-\phi/\phi_0}$.
\end{abstract}
	

\maketitle

\section{Introduction \label{sec1}}
The current accelerated expansion of the Universe was discovered by the observations of
Type Ia supernovae in 1990s \cite{Riess:1998cb,Perlmutter:1998np}, and it is now also supported by the other observations:
Cosmic Microwave Background Radiation (CMB) \cite{Komatsu:2010fb,Ade:2013zuv,Ade:2015xua},
Baryon Acoustic Oscillations (BAO)
\cite{Percival:2009xn,Blake:2011en,Beutler:2011hx,Cuesta:2015mqa,Delubac:2014aqe},
and so on.
Whereas, inflation hypothesis, which is the accelerated expansion of the Universe in
early Universe, has been broadly investigated to explain the almost isotropic signals of CMB and the flatness problem.
This acceleration cannot be described by the Einstein equations if only a usual non-relativistic matter is taken into account.
Therefore, one has to introduce a cosmological constant or dynamical scalar field(s) into general relativity, or modify the Einsteinian theory of gravity.
In this paper we treat a model which has a coupling between the kinetic term of the scalar field and the Einstein tensor
\cite{Sus:2009,SarSus:2010,Sus:2012,SusRom:2012,SkuSusTop:2013,Matsumoto:2015hua, StaSusVol, kincoupl,Germani:2010gm,GermaniKehagias:2011,Tsujikawa,Dent}.
This model can be regarded as a generalization of quintessence model. From the other hand, it can be also regarded as a special class of Horndeski theory \cite{Horndeski}, which is known as the most general single scalar field model that contains at most second derivatives in the field equations.
The dynamical properties of the model have been investigated in detail in the cases of
a power-law potential \cite{SkuSusTop:2013}, and a Higgs-like potential \cite{Matsumoto:2015hua}.

In this paper we will consider the general potential $V(\phi)$
and analyse the field equations by regarding those as an autonomous system.
The purposes of the research are not only to find some de Sitter-like solutions but also
to clarify all the attractors that the model can describe.
The contents of the paper are as follows;
in Sec.~\ref{sec2}, we derive general equations in the Friedmann-Lemaitre-Robertson-Walker (FLRW) space-time,
in Sec.~\ref{sec3}, the way to regard the field equations as an autonomous system is introduced, subsequently, the
stability of the stationary points and the properties of the other attractors are investigated in detail,
some concrete examples of the results derived in Sec.~\ref{sec3} are given in Sec.~\ref{sec4},
concluding remarks are in Sec.~\ref{sec5}.
The readers who are not so familiar with the dynamical analysis in this model
are recommended to see the figures in Sec.~\ref{sec4} before reading Sec.~\ref{sec3} because the calculations
in Sec.~\ref{sec3} are so abstract.
We use Planck units, so that $G =c =1$ in the following.
\section{Action and field equations \label{sec2}}
Let us consider the theory of gravity with the action
\begin{equation}\label{action}
S=\int d^4x\sqrt{-g}\left\{ \frac{R}{8\pi} -\big[g^{\mu\nu} + \eta G^{\mu\nu}
\big] \phi_{,\mu}\phi_{,\nu} -2V(\phi)\right\},
\end{equation}
where $V(\phi)$ is a scalar field potential, $g_{\mu\nu}$ is a metric, $R$ is the
scalar curvature, and $G_{\mu\nu}$ is the Einstein tensor.
The coupling parameter $\eta$ has the dimension of inverse mass-squared.
Note that in the literature there is discussion, which is still open, about acceptable values of $\eta$. In Refs. \cite{Germani:2010gm,GermaniKehagias:2011}
it was assumed that $\eta<0$ in order to prevent the appearance of ghosts in the model.
However, analysing scalar and tensor perturbations generated in the theories given by the action \Ref{action},
Tsujikawa in Ref. \cite{Tsujikawa} had derived more general conditions in order to avoid the appearance of scalar ghosts and Laplacian instabilities.
Generally, their fulfillment depends on particular cosmological scenarios and is not directly determined by the sign of $\eta$.
Moreover, in Ref. \cite{Dent} it was demonstrated that the necessary condition that ensures the absence of instabilities is
fulfilled for $\eta$ both positive and negative. Since the detailed investigation of the instabilities and superluminality of
the model with nonminimal kinetic coupling lies beyond the scope of the present work, hereinafter we will not impose any restrictions on values of $\eta$.

In the spatially-flat Friedmann-Robertson-Walker cosmological model the action
\Ref{action} yields the following field equations \cite{Sus:2012}
\bseq\label{genfieldeq}
\bea
\label{00cmpt}
&&3H^2=4\pi \dot{\phi}^2\left(1-9\eta H^2\right) +8\pi V(\phi),\\
&&\displaystyle
2\dot{H}+3H^2=-4\pi \dot{\phi}^2
\left[1+\eta\left(2\dot{H}+3H^2 +4H\ddot{\phi}\dot{\phi}^{-1}\right)\right]
+8\pi V(\phi),
\label{11cmpt}
\\
\label{eqmocosm}
&&\ddot\phi(1-3\eta H^2)+3H\dot\phi[1-\eta(2\dot H+3H^2)]=-V_\phi,
\eea
\eseq
where a dot denotes derivatives with respect to time, $H(t)=\dot a(t)/a(t)$ is
the Hubble parameter, $a(t)$ is the scale factor, $\phi(t)$ is a homogeneous
scalar field, and $V_\phi=dV/d\phi$.
Note that equations \Ref{11cmpt} and \Ref{eqmocosm} are of second
order, while \Ref{00cmpt} is a first-order differential
constraint for $a(t)$ and $\phi(t)$. The constraint (\ref{00cmpt}) can be
rewritten as
\beq\label{constrphigen}
\dot\phi^2=\frac{3H^2-8\pi V(\phi)}{4\pi (1-9\eta H^2)},
\eeq
or equivalently as
\beq\label{constralphagen}
H^2=\frac{4\pi \dot\phi^2+8\pi V(\phi)}{3(1+12\pi \eta\dot\phi^2)}.
\eeq
Therefore, as long as the parameter $\eta$ and the potential
$V(\phi)$ are given, the above relations provide restrictions for the
possible values of $H$ and $\dot\phi$, since they have to give rise to
non-negative $\dot\phi^2$ and $H^2$, respectively.

Hereafter we will suppose that the Hubble parameter $H$ is positive, i.e. we will take positive branch of the square root of Eq.~(\ref{constralphagen}). The case $H>0$ means that the Universe is expanding. Note that a cosmological dynamics of a contracting Universe, including ``bounce'' phenomena, was studied, for example, in \cite{hep-th:0202017,tr}.

Let us now resolve the equations \Ref{11cmpt} and \Ref{eqmocosm} with
respect to $\dot H$ and $\ddot\phi$, and then, using the relations
\Ref{constrphigen} and \Ref{constralphagen}, eliminate
$\dot\phi$ and $H$ from respective equations. As the result, we obtain
\begin{equation}\label{a2gen}
\dot H =\frac{-(1-3\eta H^2)(1-9\eta H^2)[3H^2-8\pi V(\phi)]-
	4\sqrt{\pi }\eta H \sqrt{(1-9\eta H^2)[3H^2-8\pi  V(\phi)]}\,V_\phi}
{1-9\eta H^2+54\eta^2 H^4-8\pi \eta V(\phi)(1
	+9\eta H^2)},
\end{equation}
\begin{align}\label{phi2gen}
\ddot\phi=& \big[1+12\pi \eta\dot\phi^2+96\pi^2 \eta^2\dot\phi^4
	+8\pi \eta V(\phi)(12\pi \eta\dot\phi^2-1)\big]^{-1}
        \nonumber \\
        & \times \bigg \{ -2\sqrt{3\pi}\dot\phi
	[1+8\pi \eta\dot\phi^2-8\pi \eta V(\phi)]
	\sqrt{[\dot{\phi}^2+2V(\phi)](12\pi \eta\dot\phi^2+1)}  \nonumber \\
        &-(12\pi \eta\dot\phi^2+1)(4\pi \eta\dot\phi^2+1)V_\phi \bigg \} .
\end{align}
It is seen that the $H$-equation \Ref{a2gen} in general contains
$\phi$-terms arising from the potential $V(\phi)$. At the same time, the $\phi$-equation \Ref{phi2gen} does not contains $H$-terms, so that it represents a closed second-order differential equation for the unknown function $\phi(t)$. Such the form of Eq. \Ref{phi2gen} dictates us the following strategy of solving the system of field equations \Ref{genfieldeq}. First, analyzing the $\phi$-equation \Ref{phi2gen}, we will characterize in detail a time evolution of the scalar field $\phi(t)$. Then, using the constraint \Ref{constralphagen}, we will be able to describe an evolution of the Hubble parameter $H(t)$.

\section{Scalar field as an autonomous dynamical system \label{sec3}}
In this section we will focus on solving Eq. \Ref{phi2gen} which describes a time evolution of the scalar field.

Note that Eq. \Ref{phi2gen} has a normal form, i.e. resolved with respect to the second derivative.
This allows us to use standard methods of the theory of dynamical systems.
In practice, an analysis of Eq. \Ref{phi2gen} depends on particular values of the coupling parameter $\eta$,
and hence further we will separately consider the cases $\eta>0$, $\eta<0$, and $\eta=0$.

\subsection*{The case $\eta > 0$}
Assume that $\eta$ is positive.
Now, let us introduce the set of dimensionless variables
\beq
x = \sqrt{8 \pi }\phi,\quad
y = \sqrt{8 \pi \eta}\, \dot \phi, \quad
\tau = \frac{t}{\sqrt{\eta}}, \quad
\chi(x) = 8 \pi  \eta V(x/\sqrt{8 \pi }).
\label{dlp}
\eeq
Here $\chi(x)$ is a dimensionless potential, and as well as $V(\phi)$, it is  non-negative, i.e. $\chi(x)\geq0$.  By using dimensionless variables, we can rewrite Eq. \Ref{phi2gen} as the following autonomous dynamical system:
\begin{align}
\frac{dx}{d \tau} &= y,
\label{A1} \\
\frac{dy}{d \tau} &
= \frac{-  \sqrt{3} y \left [ 1+y^2- \chi (x) \right ] \sqrt{[{\textstyle\frac{1}{2}}y^2+\chi (x)]\left({\textstyle\frac{3}{2}}y^2+1 \right)}
-  \left ( \frac{3}{2}y^2 +1 \right ) \left ( \frac{1}{2}y^2 +1 \right ) \chi ' (x)}
{1+\frac{3}{2}y^2+\frac{3}{2}y^4+ \chi (x) \left ( \frac{3}{2}y^2 -1 \right )}
\label{A2}.
\end{align}
As seen in the system \Ref{A1}-\Ref{A2}, the degrees of freedom of the model, which are
the function $V(\phi)$ and the constant $\eta$ in Eq.~\Ref{phi2gen}, are simply expressed as $\chi (x)$.
Therefore, the mathematical properties of Eq.~\Ref{A1}-\Ref{A2} only depend on $\chi (x)$,
although the physical properties of the original equation \Ref{phi2gen} depend on both of $V(\phi)$ and $\eta$.
Below we examine in detail basic features of the system (\ref{A1})-(\ref{A2}).

\subsubsection{Stationary points}
Stationary points $(x_0,y_0)$ are those where $dx/d\tau=0$ and $dy/d\tau=0$.
Hence, from Eq. \Ref{A1} it follows that any stationary point of the system (\ref{A1})-(\ref{A2}) has $y_0=0$. Then, Eq. \Ref{A2} yields
\beq
\left.\frac{dy}{d \tau}\right|_{(x,y)=(x_0,0)} =\frac{- \chi'(x_0)}{1-\chi (x_0)}.
\eeq
The equality $(dy/d\tau)|_{(x_0,0)}=0$ is fulfilled if $\chi'(x_0)=0$ and $\chi(x_0)\neq 1$.

A stability of stationary points with respect to small perturbations $\delta x$ and $\delta y$ can be investigated by fluctuating Eqs.~(\ref{A1})-(\ref{A2}) as
\begin{equation}
\frac{d}{d \tau}
\left(
\begin{array}{c}
\delta x  \\
\delta y
\end{array}
\right)
\Bigg \vert _{(x,y)=(x_0,0)}
=
\left(
\begin{array}{cc}
0 & 1 \\
\frac{\partial y'_\tau}{\partial x} & \frac{\partial y'_\tau}{\partial y}
\end{array}
\right)
\Bigg \vert  _{(x,y)=(x_0,0)}
\left(
\begin{array}{c}
\delta x  \\
\delta y
\end{array}
\right) ,
\label{PA}
\end{equation}
where $y'_\tau=dy/d\tau$.
A character of stationary points $(x_0,0)$ and a behavior of phase trajectories in their vicinity are determined by the eigenvalues of the matrix standing in the right-hand side of Eq.~(\ref{PA}).
In particular, $(x_0,y_0)$ is stable if all real parts of eigenvalues are negative.

One can easily check that a $2\times 2$ matrix $\{(0,a),(b,c) \}$ has, generally speaking, two eigenvalues $(c \pm \sqrt{c^2 + 4ab})/2$. Both of them have negative real parts if and only if $c<0$ and $ab<0$. Moreover, since a configuration of phase trajectories is determined by imaginary parts of eigenvalues, phase trajectories form spirals if $c^2+4ab<0$, and straight lines if $c^2+4ab \geq 0$. Using these algebraic facts, we can conclude that, in our case, the stationary points $(x_0,y_0)$ are stable if and only if
\begin{align}
\frac{\partial y'_\tau}{\partial x} \bigg \vert _{(x,y)=(x_0,0)} <0 \qquad
\mathrm{and} \qquad
\frac{\partial y'_\tau}{\partial y} \bigg \vert _{(x,y)=(x_0,0)} <0, \label{stability_cond}
\end{align}
so that from Eqs. \Ref{A1}-\Ref{A2} we respectively find
\begin{align}
\frac{- \chi ''(x_0)}{1- \chi (x_0)} <0 \qquad \label{stability_cond2}
\mathrm{and} \qquad
- \sqrt{3 \chi (x_0)} <0.
\end{align}
The second inequality of \Ref{stability_cond2} is always satisfied if only $\chi (x_0)>0$, whereas a fulfillment of the first inequality depends on the value of $\eta$.
Note that in case $\chi (x_0)=0$ we need in more delicate investigation of stability conditions, because the second equation of \Ref{stability_cond2} vanishes.

If real parts of the eigenvalues are equal to zero, it means that stability
cannot be understood only from linear perturbation terms.
Now, next leading order terms, i.e. second order terms should be taken into account.
Expanding Eqs.~(\ref{A1}) and (\ref{A2}) around $(x,y)=(x_0,0)$ by applying
$x\to x_0+\delta x$ and $y\to 0+\delta y$ yields
\begin{align}
\frac{d \delta x}{d \tau} &= \delta y,
\label{A1cwx00} \\
\frac{d \delta y}{d \tau} &\textstyle
= - \chi '' (x_0)\delta x - \frac{1}{2} \chi '''(x_0) \delta x^2 - \sqrt{\frac{3}{2}\left [\delta y^2 + \chi ''(x_0)\delta x^2 \right ]}\delta y
+ O(\delta ^3)
\label{A2cwx00},
\end{align}
where $\delta \sim \delta x \sim \delta y$.
Taking into account the definition \Ref{dlp} for dimensionless variables, one can see that Eqs.~(\ref{A1cwx00}) and (\ref{A2cwx00}) do not contain the coupling parameter $\eta$. This means that these equations are equivalent to those in case of $\eta=0$, i.e. without kinetic coupling.
An approximate solution of Eqs.~(\ref{A1cwx00}) and (\ref{A2cwx00}) can be easily found by the WKB method \cite{Star}. In the dimensional quantities the solution reads
\begin{equation}
\delta\phi =\frac{1}{\sqrt{3\pi}} \frac{\sin\big(\sqrt{V''(\phi_0)}\, t\big)}{\sqrt{V''(\phi_0)}\, t}.
\end{equation}
Therefore, the stationary point is stable if $V''(\phi_0)>0$ and unstable if $V''(\phi_0)<0$.

\subsubsection{Limit cycles}
Periodic solutions or, equivalently, limit cycles in a phase space play a peculiar role. In cosmology they might represent cyclic scenarios of the Universe evolution. The Bendixson-Dulac theorem on dynamical systems states the non-existence conditions of periodic solutions on a phase plane (see, for example, \cite{BDtheorem}).
Namely, it states that if there exists a $C^1$ function $\varphi(x,y)$ (called the Dulac function) such that the expression
\begin{equation}
\Phi(x,y)= \frac{\partial(\varphi f)}{\partial x}+
\frac{\partial(\varphi g)}{\partial y}
\label{BD}
\end{equation}
has the same sign ($\neq 0$) almost everywhere in a simply connected region of the plane, then the plane autonomous system
$$
\frac {dx}{d \tau}=f(x,y),\quad \frac {dy}{d \tau}=g(x,y),
$$
has no periodic solutions lying entirely within the region.
``Almost everywhere'' means everywhere except possibly in a set of measure 0,
such as a point or line.

Assuming $\chi(x)=0$ and substituting the functions $f(x,y)$ and $g(x,y)$ from the autonomous system \Ref{A1}-\Ref{A2} into the expression \Ref{BD}, we obtain
\begin{equation}
\Phi(x,y)=\varphi'_x(x,y)y +\varphi'_y(x,y) y'_\tau -\varphi(x,y) \frac{\sqrt{3}y^2 (8+34y^2+39y^4+18y^6+9y^8)}
{\sqrt{y^2(2+3y^2)}(2+3y^2+3y^4)^2}.
\end{equation}
Now, the choice $\varphi(x,y)\equiv C$, where $C$ is a positive constant, guarantees that $\Phi(x,y)<0$ everywhere except $y=0$.
Therefore, according to the Bendixson-Dulac theorem, in the cosmological model with $\chi(x)=0$ there exist no periodic solutions.
The consideration of limit cycles in the case of an arbitrary function $\chi(x)$ lies beyond the scope of this paper. However, it is worth noticing that the particular examples of $\chi(x)$ considered below do not contain any periodic solutions (see Figs. \ref{sr}-\ref{exp}).
Note also that the role of limit cycles in cosmology was well studied in \cite{LC1,LC2}.

\subsubsection{Nullclines $y'_\tau = 0$}
The relation $y'_\tau = 0$ determines on the phase plane a curve named a nullcline.
From Eq. \Ref{A2} we obtain the following equation for the nullcline:
\begin{equation}
\sqrt{3} y \left[1+y^2-\chi(x)\right]
\sqrt{[{\textstyle\frac{1}{2}}y^2+\chi (x)]\left({\textstyle\frac{3}{2}}y^2+1 \right)}
+\left(\textstyle\frac{3}{2}y^2 +1 \right)
 \left(\textstyle\frac{1}{2}y^2 +1 \right) \chi'(x)=0.
\label{110}
\end{equation}
Generally, this equation cannot be resolved with respect to $y$,
however, we can study its asymptotical properties.

Let us consider the limit $x\rightarrow \pm \infty$. First, suppose that in this case the potential $\chi(x)$ tends to a finite constant value, i.e. $\chi(x)\to\chi_\pm$. Then, respectively, $\chi'(x)\to 0$, and Eq.~\Ref{110} has the following asymptotical solutions: $y=0$, $y=\sqrt{\chi_\pm-1}$, and $y=-\sqrt{\chi_\pm-1}$ with $\chi_\pm>1$.

In case $\chi (x)\to\infty$ at $x\rightarrow \pm \infty$, one can obtain several various asymptotes.
First, let us assume that $y^2 \ll \chi (x)$ and $y^2\gg 1$, then from Eq.~\Ref{110} we find the nullcline
\begin{equation}
\mathop{\rm sgn}(y)\, y^2 = 2 \sqrt{2}\, \frac{\chi ^{3/2}(x)}{\chi '(x)},
\label{as1}
\end{equation}
where a sing of $y$ should coincide with that of $\chi'(x)$ and the relation $|\chi'(x)|>\sqrt{\chi(x)}$ has to be fulfilled.

If $y^2 \ll \chi (x)$ and $y^2=O(1)$,
an asymptote of the  nullcline takes the form
\begin{equation}\textstyle
\sqrt{3\chi ^3}y - \sqrt{\frac{3}{2}y^2+1}\left ( \frac{1}{2}y^2 +1 \right ) \chi ' =0
\label{as2},
\end{equation}
with the relation $\chi^3(x)>\chi'^2(x)$.
In particular,
if the condition $\chi^3(x)\gg \chi'^2(x)$ holds, then we have
\begin{equation}
y^2 = \frac{\chi ^{\prime 2}(x)}{3 \chi ^3 (x)}.
\label{as3}
\end{equation}

If $y^2\gg 1$ and $y^2/\chi (x)=O(1)$, the nullcline exists only if
$\chi'^2(x) \lesssim \chi(x)$,
and is expressed as
\begin{equation}
\textstyle
(y^2 -\chi (x)) \sqrt{\frac{1}{2} y^2 \left ( \frac{1}{2} y^2 + \chi (x) \right )} + \frac{1}{4} y^3 \chi '(x) =0.
\label{sas}
\end{equation}
In particular, the nullcline \Ref{sas} is simplified for the case $\chi'^2(x) \ll \chi(x)$ as
\begin{equation}
y= \pm \sqrt{\chi (x)}.
\label{sas2}
\end{equation}

If $y^2\gg1$ and $y^2 \gg \chi(x)$, we obtain the nullcline
\begin{equation}
y = - \frac{1}{2} \chi'(x).
\label{as4}
\end{equation}

All nullclines derived in this paragraph are listed in the table \ref{Table1}.
\begin{table}[t]
	\begin{tabular}{|c||c|c|c|c||c|} \hline
		Region & Range of $y$ & $\chi (x \rightarrow \pm \infty)$ & Sign of $y \chi '$ & Existence conditions & Nullclines \\ \hline \hline
		$y^2 \gg \chi$ & $\vert y \vert \gg 1$ & infinite & $-$ & $\vert \chi ' \vert \gg \chi ^{1/2}$ & $y=- \chi ' /2$ \\ \hline
		& $\vert y \vert \gg 1$ & infinite & $\pm$ & $\vert \chi ' \vert \sim \chi ^{1/2}$ & Eq.~\Ref{sas} \\ \cline{3-6}
		$y^2 \sim \chi$ &  & infinite & $\pm$ & $\vert \chi ' \vert \ll \chi ^{1/2}$ & $y=\pm \sqrt{\chi }$ \\ \cline{2-6}
		& $y = O(1)$ & finite & $0$ & $ \chi >1$ & $y= \pm \sqrt{\chi -1}$ \\ \hline
		& $\vert y \vert \gg 1$ & infinite & $+$ & $\chi ^{1/2} \ll \vert \chi ' \vert \ll \chi ^{3/2}$ & $y^3/\vert y \vert = 2\sqrt{2 \chi ^3}/ \chi ' $ \\ \cline{2-6}
		$y^2 \ll \chi$ & $ y = O(1)$ & infinite & $+$ & $\vert \chi ' \vert \sim \chi ^{3/2}$ & Eq.~\Ref{as2} \\ \cline{2-6}
		& $\vert y \vert \ll 1$ & infinite & $+$ & $\vert \chi ' \vert \ll \chi ^{3/2}$ & $y= \chi ' /\sqrt{3 \chi ^3}$ \\ \cline{2-6}
		& $\vert y \vert \ll 1$ & finite & $0$ & $\chi \neq 1$ & $y= -\chi '/[\sqrt{3 \chi}(1-\chi)]$ \\ \hline
	\end{tabular}
	\caption{Nullclines in the region $\vert x \vert \gg 1$ or $\vert y \vert \gg 1$.
		The signature of $y \chi '$ implies that the existence regions
		of the nullclines are constrained. For example, the continuous functions $\chi (x) >0$ and
		$\chi ' (x)$ cannot satisfy the conditions
		$\chi (+ \infty) \gg 1$ and $y \chi '(+ \infty) >0$ in the fourth quadrant.
		Therefore, the nullclines of $y \chi '>0$ only exist in the first and third quadrants.
	}
	\label{Table1}
\end{table}

\subsubsection{Critical curves $y'_\tau = \pm \infty$}
The other notable curves on the phase plane are those where $y'_\tau = \pm \infty$.
Such the curves are dubbed as {\em critical}.
The instability of $\dot H$, in other words, the curvature singularity arise on the curve.
From Eq. \Ref{110} one can see that
sufficient conditions providing $y'_\tau = \pm \infty$ are $\chi (x) = \pm \infty$, or $\chi'(x) = \pm \infty$,
or
\begin{equation}
1+\frac{3}{2}y^2+\frac{3}{2}y^4+ \chi (x) \left ( \frac{3}{2}y^2 -1 \right )=0.
\label{100}
\end{equation}

	

For positive $\chi(x)$ the equation \Ref{100} could be satisfied if only $\vert y \vert < \sqrt{2/3}$. If $\chi (x)$ tends to infinity in the limit $x \rightarrow \pm \infty$,
then the critical curve \Ref{100} asymptotically approaches $y= \pm \sqrt{2/3}$.
Resolving Eq.~(\ref{100}) with respect to $\chi (x)$ gives a constraint $\chi (x) \geq 1$. This result means that if $\eta$ is small enough to provide $8 \pi \eta V(\phi) < 1$ for all $\phi$, then there exists no critical curves on the phase plane
and if $\eta$ is large enough to provide $8\pi\eta V(\phi)\geq 1$ for some $\phi$, then the critical curves appear.

It is worth noticing that the critical curves \Ref{100} divide the phase plane
into several isolated regions.
A phase trajectory running through the isolated region cannot cross the critical curve where $y'_\tau = \pm \infty$, and hence one faces with singularities or indefinableness of physical quantities.
Note that a study of such singular curves and the behavior of trajectories around them
was done in \cite{LC2,CC}
for $k$-essence-like theories.


\subsubsection{Behavior of phase trajectories in regions separated by critical curves}
\begin{table}[t]
  \begin{tabular}{|c|c||c|c||c|} \hline
    \multicolumn{2}{|c||}{Type of the function $\chi (x)$} & Asymptotes & Existence conditions & Hubble rate \\ \hline \hline
    \multicolumn{2}{|c||}{} & $y= -\chi '/[\sqrt{3 \chi}(1-\chi)]$ & $\chi (\pm \infty) \neq 1$, & $H=\sqrt{\frac{8 \pi V}{3}}$ \\
    \multicolumn{2}{|c||}{$\chi (\pm \infty)$ is finite} & & $\chi ' /(1- \chi)<  0$ &  \\ \cline{3-5}
    \multicolumn{2}{|c||}{} & $y= \pm \sqrt{\chi -1}$ & $\chi (\pm \infty) > 1$ & $H=\frac{1}{\sqrt{3 \eta}}$ \\ \hline
    & $\vert \chi ' \vert \ll \chi ^{1/2}$ & $y= \frac{\chi '}{\sqrt{3 \chi ^3}}$ & & $H=\sqrt{\frac{8 \pi V}{3}}$ \\ \cline{3-5}
    & & $y= \pm \sqrt{\chi} $ &  & $H=\frac{1}{\sqrt{3 \eta}}$ \\ \cline{2-5}
    & $\vert \chi ' \vert \sim \chi ^{1/2}$ & $y= \frac{\chi '}{\sqrt{3 \chi ^3}}$ & & $H=\sqrt{\frac{8 \pi V}{3}}$ \\ \cline{3-5}
    & & $y= s \sqrt{ \chi} $, $s \neq 0, \pm 1$ &  & $H=\sqrt{\frac{2+s ^2}{9 s ^2 \eta}}$ \\ \cline{2-5}
    $\chi (\pm \infty)$ is infinite & $\chi ^{1/2} \ll \vert \chi ' \vert \ll \chi ^{3/2}$ & $y= \frac{\chi '}{\sqrt{3 \chi ^3}}$ & & $H=\sqrt{\frac{8 \pi V}{3}}$ \\ \cline{3-5}
    & & $y^2=u \frac{\sqrt{\chi ^3}}{\chi '}$, $u \neq 0, \pm 2\sqrt{2}$ & & $H=\sqrt{\frac{2V_\phi }{9 u (\kappa V)^{1/2}}}$ \\ \cline{2-5}
    & & $y=c_1, 0<\vert c_1 \vert < \sqrt{\frac{2}{3}}$ & $0< \vert \frac{\chi '}{\chi ^{3/2}} \vert < \frac{3}{4}$ & $H=\sqrt{\frac{16 \pi V}{6+9c_1^2}}$ \\ \cline{3-5}
    & $\vert \chi ' \vert \sim \chi ^{3/2}$ & $y=c_2, \sqrt{\frac{2}{3}}<\vert c_2 \vert $ & $0< \vert \frac{\chi '}{\chi ^{3/2}} \vert < \frac{3}{4}$ & $H=\sqrt{\frac{16 \pi V}{6+9c_2^2}}$ \\ \cline{3-5}
    & & $y=\pm \sqrt{\frac{2}{3}}$ & $\frac{3}{4}< \vert \frac{\chi '}{\chi ^{3/2}} \vert$ & $H=\sqrt{\frac{4 \pi V}{3}}$ \\ \cline{2-5}
    & $\chi ^{3/2} \ll \vert \chi ' \vert$ & $y=\pm \sqrt{\frac{2}{3}}$ & & $H=\sqrt{\frac{4 \pi V}{3}}$ \\ \hline
  \end{tabular}
  \caption{Asymptotes of the trajectories in the region $\vert x \vert \gg 1$. }
  \label{table2}
\end{table}

First, we can find that there are no attractors in the region $\chi ^{1/2}, |\chi'| \ll |y|$ and $|y|\gg 1$,
because the signature of $dy/d \tau$ is opposite to that of $y$.
It can be also seen that the attractors in the limit $x \rightarrow \pm \infty$ or $y \rightarrow \pm \infty$
are only located in the region $x \rightarrow + \infty$ and $y >0$ and the region $x \rightarrow - \infty$ and $y <0$
by taking into account the signature of $dx/ d \tau$.
In the following, we will investigate the behaviors of the trajectories in the regions
$x \gg + 1$, $y >0$ and $x \ll -1$, $y <0$ by using case analysis
with respect to the form of the function $\chi (x)$ in the limit $x \rightarrow \pm \infty$.

\paragraph{The case $\chi (\pm \infty)$ is finite.}
When the function $\chi (x)$ is finite in the limit $x \rightarrow \pm \infty$,
there are only nullclines $y_{n\pm} = \pm \sqrt{ \chi -1}$ and $y= -\chi '/[\sqrt{3 \chi}(1-\chi)]$
in the region $\vert x \vert \gg 1$ or $\vert y \vert \gg 1$
if we assume a continuity of $\chi (x)$ and $\chi ' (x)$
as shown in the last subsection.
The critical curves \Ref{100} are written down as
\begin{equation}
y_{s\pm} = \pm \sqrt{-\frac{1+ \chi}{2} + \frac{1}{2}\sqrt{(1+\chi)^2 -\frac{8}{3}(1-\chi)}}
\label{crif}
\end{equation}
in that region. The existence conditions for the nullclines $y_{n\pm} = \pm \sqrt{ \chi -1}$ and
it for the critical curves \Ref{crif} are same, i.e. $\chi >1$.
The absolute value of $y_{n\pm}$ is always larger than that of $y_{s\pm}$, so
the signature of $dy/d \tau$ in the case $y>0$ is determined as follows;
$dy/d \tau $ is negative above $y_{n+} = \sqrt{ \chi -1}$, it becomes positive in the region $y_{s+}<y<y_{n+}$,
and it becomes negative if $-\chi '/[\sqrt{3 \chi}(1-\chi)]<y<y_{s+}$.
By taking into account the signatures of $dy/d \tau$ and $dx/ d \tau$,
we can find the nullcline $y_{n +} = + \sqrt{ \chi -1}$ is an attractor in the limit $x \rightarrow + \infty$,
and $y_{n -} = - \sqrt{ \chi -1}$ is an attractor in the limit $x \rightarrow - \infty$.
The stability of the curve $y= -\chi '/[\sqrt{3 \chi}(1-\chi)]$ is determined by the sign of
$\chi ' /(1- \chi)$ because it determine the sign of $y$ in the region $\vert x \vert \gg 1$.
The condition that the curve should be in the first or third quadrant yields
the stability condition $\chi ' /(1-\chi) <  0$.
The value of the Hubble rate function on $y_{n\pm} = \pm \sqrt{ \chi -1}$ and $y=0$ are given as $H=1/ \sqrt {3 \eta}$ and
$H= \sqrt{8 \pi V/3}$, respectively.
\paragraph{The case $\chi (\pm \infty)$ is infinite and $\vert \chi ' (\pm \infty) \vert \ll \chi ^{1/2}(\pm \infty)$.}
If the functions $\chi (x)$ and $\chi ' (x)$ are continuous functions, then
$\chi ' (+ \infty) >0$ ($\chi ' (- \infty) <0$) holds when $\chi (+ \infty ) \rightarrow + \infty $ ($\chi (- \infty ) \rightarrow + \infty $) is satisfied.
Therefore, we will consider the cases $\chi ' (+ \infty) >0$ and $\chi ' (- \infty) <0$ in the following.
If $\chi (\pm \infty)$ is infinite and $\vert \chi ' (\pm \infty) \vert$ is much less than $\chi ^{1/2}(\pm \infty)$, there are two kinds of
nullclines in the limit $x \rightarrow \pm \infty$.
The nullcline $y = \pm \sqrt{\chi (x)}$ is located in the region $y^2 \sim \chi (x)$ and
the nullcline $y = \chi'(x) / \sqrt{3 \chi ^3(x)}$ is located in the region $y^2 \ll \chi (x)$.
On the other hand, the critical curves \Ref{100} are approximately written down as $y = \pm \sqrt{2/3}$ in the region
$\vert x \vert \gg 1$.
In the right half of the x-y phase diagram, these critical curves are located in order from the top as $y = \sqrt{\chi (x)}$,
$y = \sqrt{2/3}$, $y = \chi'(x) / \sqrt{3 \chi ^3(x)}$, $y = -\sqrt{2/3}$, and $y = -\sqrt{\chi (x)}$.
The signature of $dy/d \tau$ is $-$ in the range $y>\sqrt{\chi (x)}$, is $+$ in $\sqrt{2/3}<y<\sqrt{\chi (x)}$,
is $-$ in $\chi'(x) / \sqrt{3 \chi ^3(x)}<y<\sqrt{2/3}$, and so on.

By taking that $\chi'(x) / \sqrt{3 \chi ^3(x)} >0$ and $\chi'(x) / \sqrt{3 \chi ^3(x)} \simeq const. $ hold
in the region $\vert x \vert \gg 1$ into account,
it is seen that $y=\chi'(x) / \sqrt{3 \chi ^3(x)}$ is an asymptote of the trajectories.
The Hubble rate function on this asymptote is given by $H = \sqrt{8 \pi V/3}$.
Whereas, the nullcline $y= \sqrt{\chi}$ is also an asymptote of the trajectories in the first quadrant and in the third quadrant as seen in the following.
The inclination of $y$ with respect to $x$ of the nullcline $y= \pm \sqrt{\chi}$ is given as
$dy/dx = \chi '/(2 \sqrt{\chi}) \rightarrow 0$, whereas, the inclination of the
trajectories on the curve $y= s \sqrt{\chi}$ is
$(dy/d\tau)/(dx/d\tau) = 2(1- s ^2)\sqrt{s ^2(s ^2 +2)}/(3 s ^5)$.
Therefore, if the trajectories are once on the nullcline $y= \pm \sqrt{\chi}$, the trajectories continue being on it.
At the same time, the condition $\vert (dy/d\tau)/(dx/d\tau) \vert \gtrsim 1$ is satisfied in the adjoining regions
$1 \ll \vert y \vert \ll \sqrt{\chi}$ and $\sqrt{\chi} \ll \vert y \vert$,
so the trajectories that are far from $y= \pm \sqrt{\chi}$ will approach the nullcline $y= \pm \sqrt{\chi}$.
The Hubble rate function on this nullcline is given by $H = 1/\sqrt{3 \eta}$.
\paragraph{The case $\chi (\pm \infty)$ is infinite and $\vert \chi ' (\pm \infty) \vert \sim \chi ^{1/2} (\pm \infty)$.}
The behavior of the trajectories in the case that $\chi (\pm \infty)$ is infinite and $\vert \chi ' (\pm \infty) \vert \sim \chi ^{1/2} (\pm \infty)$
is almost same as that we considered in the last paragraph. However, we should
take care of the behaviors of the trajectories in the region $y^2 \sim \chi (x)$.
In general, the nullcline in the region $y^2 \sim \chi (x)$ is obtained by solving Eq.~\Ref{sas}.
If we only evaluate the leading part of the terms, we can treat $\chi '$ as
$\chi ' \propto \chi ^{1/2}$ because $\vert \chi ' (\pm \infty) \vert \sim \chi ^{1/2} (\pm \infty)$.
Then, we have $y=u \chi ^{1/2}$ as a solution of Eq.~\Ref{sas}, where $u \neq 0,1$ is a real number.
There is only one value of $u$ that can satisfy Eq.~\Ref{sas} if we treat $\chi '$ as $\chi ' = \beta \chi ^{1/2}$, $\beta \in \mathbb{R}_{\neq 0}$.
On the nullclines, $(dy/d\tau )/(dx/d \tau) =0$ is satisfied because $dy/d\tau =0$, whereas,
the derivative of the nullcline $y=u \chi ^{1/2}$ yields $dy/dx \neq 0$.
Therefore, the nullcline $y=u \chi ^{1/2}$ is not an asymptote of the trajectories.
At the same time, if we consider the curve $ y = f(x)=\gamma \chi ^{1/2}$, where $\gamma \neq 0,1, u$ is a real number,
we have $(dy/d\tau )/(dx/d \tau)\vert_{y=f(x)} = O(1)$ and $df(x)/dx= O(1)$.
Therefore, we can find the asymptotes by solving the equation $(dy/d\tau )/(dx/d \tau)\vert_{y=f(x)} =df(x)/dx$ with respect to $\gamma$.
There is only one value of $\gamma$ that can satisfy this equality.
The Hubble rate function on the asymptote is expressed as $H = \sqrt{(2+ \gamma ^2)/(9 \gamma ^2 \eta)}$.

\paragraph{The case $\chi (\pm \infty)$ is infinite and $\chi ^{1/2} (\pm \infty) \ll \vert \chi ' (\pm \infty) \vert \ll \chi ^{3/2} (\pm \infty)$.}
If $\chi (\pm \infty)$ is infinite and $\chi ^{1/2} (\pm \infty) \ll \vert \chi ' (\pm \infty) \vert \ll \chi ^{3/2} (\pm \infty)$,
there are three kinds of nullclines; $y= \chi '/\sqrt{3 \chi ^3}$, $y^3/\vert y \vert = 2 \sqrt{2 \chi ^3}/\chi '$,
and $y= - \chi '/2$.
The behavior of the trajectories around the first one has been already considered in the other cases, and it is not changed in this case.
The second one is located in the region $y^2 \ll \chi$ if $y \chi ' >0$, while, the third one is
located in the region $y^2 \gg \chi$ if $y \chi ' <0$.
In the region $2/3< y^2 \ll 2 \sqrt{2 \chi ^3}/\vert \chi ' \vert$, the condition $(dy/d \tau)/(dx /d \tau) \gg 1$ is always satisfied,
moreover, this inclination is more than that of the nullclines $y^3/\vert y \vert = 2 \sqrt{2 \chi ^3}/\chi '$,
so the trajectories in this region approach the nullclines $y^3/\vert y \vert = 2 \sqrt{2 \chi ^3}/\chi '$.

In the region $y^2 \sim \chi$, the direction of the trajectories depends on the signature of $y \chi '$ as
$(dy/d\tau )/(dx/d\tau ) \sim - \chi ' /y$.
If $y \chi ' >0$, there are only nullclines $y^3/\vert y \vert = 2 \sqrt{2 \chi ^3}/\chi '$
for $y^2 > 2/3$ and the inclination of the trajectories is negative in the region $y^2 \sim \chi$,
so the curves $y^3/\vert y \vert = 2 \sqrt{2 \chi ^3}/\chi '$ look stable when viewed from a distance.
However, they are not attractors because the inclement of them do not vanish although $dy/d \tau =0$ is satisfied on them.
The attractors are, in fact, given by $y^2 = c_1 \sqrt{\chi ^3}/\chi '$, where $c_1 \neq \pm 2 \sqrt{2}$ is a real number.
The value of $c_1$ is exactly determined if the index $i$ and the coefficient $c$ of $\chi ' (\pm \infty) = c \chi ^{i} (\pm \infty)$ is given.
The difference between the nullclines and the attractor is only the coefficient of the terms.
On the other hand, if $y \chi ' <0$, there are only nullclines $y= - \chi '/2$ for $y^2 > 2/3$,
however, they are only located in the second quadrant and in the fourth quadrant if $\chi$ is positive.
Therefore, the nullclines $y= - \chi '/2$ have nothing to do with the attractors in $x \rightarrow \pm \infty$.

\paragraph{The case $\chi (\pm \infty)$ is infinite and $\vert \chi ' (\pm \infty) \vert \sim \chi ^{3/2} (\pm \infty)$.}
If $\chi (\pm \infty)$ is infinite and $\vert \chi ' (\pm \infty) \vert \sim \chi ^{3/2} (\pm \infty)$,
there are two kinds of nullclines; $y=- \chi ' /2$ and Eq.~(\ref{as2}).
The nullclines $y=- \chi ' /2$ have nothing to do with the attractors in the region $x \rightarrow \pm \infty$,
whereas, Eq.~(\ref{as2}) gives the characteristic configuration of the attractors.
By assuming the function $\chi ' (x)$ as $\chi ' (x) = k \chi ^{3/2}(x)$
in the limit $x \rightarrow \pm \infty$, Eq.~(\ref{as2}) gives
\begin{equation}
k = \frac{\sqrt{3}y}{\sqrt{\frac{3}{2}y^2+1}\left ( \frac{1}{2}y^2 +1 \right )}. \label{325}
\end{equation}
If $0< \vert k \vert < 3/4$, then there are two solutions in Eq.~(\ref{325}) and the range of the
solutions are given by $0< \vert y \vert < \sqrt{2/3}$ and $\vert y \vert > \sqrt{2/3}$, respectively.
If $k = \pm 3/4$, then
there is one solution $y= \pm \sqrt{2/3}$, else if $\vert k \vert >  3/4$, then there is no real solutions.
Taking the existence of the critical curves $y = \pm \sqrt{2/3}$ into account
gives the existence condition for the nullclines $y= const.$ as $0< \vert k \vert < 3/4$.
The nullclines $y= const.$ are always stable for $y$ direction, however, are only stable
in the case $\chi '(+ \infty) >0$ or $\chi '(- \infty) <0$ for $x$ direction, therefore, they are the attractors
if they are in the first quadrant or in the third quadrant.
The case $ \vert k \vert > 3/4$ is same as the following case; there are no nullclines in the first and the third quadrant and the
critical lines $y = \pm \sqrt{2/3}$ become stable.

\paragraph{The case $\chi (\pm \infty)$ is infinite and $\chi ^{3/2} (\pm \infty) \ll \vert \chi ' (\pm \infty) \vert $.}
There are only the nullclines $y= \chi ' /2$ in the limit $x \rightarrow \pm \infty$
in the case $\chi (\pm \infty)$ is infinite and $\chi ^{3/2} (\pm \infty) \ll \vert \chi ' (\pm \infty) \vert $.
Therefore, the critical curves that have a relation with the attractors are only $y = \pm \sqrt{2/3}$.
In fact,  $y = \sqrt{2/3}$ is the attractor in the limit $x \rightarrow + \infty$ and $y = -\sqrt{2/3}$ is the
attractor in the limit $x \rightarrow - \infty$, although $dy/d \tau$ becomes singular on the curves $y = \pm \sqrt{2/3}$.

The asymptotes in the region $\vert x \vert \gg 1$ of the case $\eta > 0$ are summarized in Table~\ref{table2}.

\subsection*{The case $\eta < 0$}
\begin{table}[t]\label{tablen}
  \begin{tabular}{|c|c||c|c||c|} \hline
    \multicolumn{2}{|c||}{Type of the function $\chi (x)$}& $(x,y)$ of the attractors & Stability conditions & Hubble rate \\ \hline \hline
    \multicolumn{2}{|c||}{$\chi (\pm \infty)$ is finite} & $(x_0 ,0)$, s.t. $\chi '(x_0) =0$ & $\chi '' (x_0)  >0$ & $H=\sqrt{\frac{8 \pi V}{3}}$ \\ \cline{3-5}
    \multicolumn{2}{|c||}{} & $(\pm \infty , 0)$ & $\chi '(\pm \infty) = \mp 0$ & $H=\sqrt{\frac{8 \pi V}{3}}$ \\ \hline
    & $\vert \chi ' \vert \ll \chi ^{3/2}$ & $(x_0 ,0)$, s.t. $\chi '(x_0) =0$ & $\chi '' (x_0)  >0$ & $H=\sqrt{\frac{8 \pi V}{3}}$ \\ \cline{2-5}
    & $\vert \chi ' \vert \sim \chi ^{3/2}$ & $(x_0 ,0)$, s.t. $\chi '(x_0) =0$ & $\chi '' (x_0)  >0$ & $H=\sqrt{\frac{8 \pi V}{3}}$ \\ \cline{3-5}
    $\chi (\pm \infty)$ is infinite & & $(- \infty, y_0)$, $-\sqrt{\frac{2}{3}} <y_0 <0$ & $\chi (- \infty) \rightarrow + \infty$ & $H=\sqrt{\frac{16 \pi  V}{3(2-3y_0^2 )}}$ \\ \cline{2-5}
    & $ \chi ^{3/2} \ll \vert \chi ' \vert$ & $(x_0 ,0)$, s.t. $\chi '(x_0) =0$ & $\chi '' (x_0)  >0$ & $H=\sqrt{\frac{8 \pi V}{3}}$ \\ \cline{3-5}
    & & $(- \infty , -\sqrt{\frac{2}{3}})$ & $\chi (- \infty) \rightarrow + \infty$ & $H=+ \infty$ \\ \hline
  \end{tabular}
  \caption{Attractors in the case $\eta < 0$. }
  \label{table3}
\end{table}

In case $\eta$ is negative, we need to redefine the set of dimensionless variables as follows
\beq
x=\sqrt{8\pi}\phi,\quad
y=\sqrt{8\pi |\eta|}\, \dot \phi, \quad
\tau=\frac{t}{\sqrt{|\eta|}}, \quad
\chi(x)=8\pi |\eta| V(x/\sqrt{8\pi}).
\label{dln}
\eeq
Since $V(\phi)$ is assumed to be non-negative, we have $\chi(x)\geq 0$.
Now, we can rewrite Eq. \Ref{phi2gen} as the following autonomous dynamical system:
\begin{align}
\frac{dx}{d \tau} &=y,
\label{A1n} \\
\frac{dy}{d \tau} &
=\frac{-\sqrt{3} y [1-y^2+\chi(x)] \sqrt{[{\textstyle\frac{1}{2}}y^2+\chi (x)]\left(1-{\textstyle\frac{3}{2}}y^2 \right)}
- \left (1- \frac{3}{2}y^2  \right ) \left (1- \frac{1}{2}y^2  \right ) \chi ' (x)}
{1-\frac{3}{2}y^2+\frac{3}{2}y^4+ \chi (x) \left ( \frac{3}{2}y^2 +1 \right )}
\label{A2n}.
\end{align}
The argument under the square root should be non-negative, hence we have $\left(1-{\textstyle\frac{3}{2}}y^2 \right)\geq 0$. Because of this condition the denominator in \Ref{A2n} is strictly positive and cannot vanish unlike the case $\eta >0$.

From (\ref{A1n})-(\ref{A2n}) one can easily find that a stationary point $(x_0,y_0)$ is determined by the following conditions:
\begin{equation}
y_0=0, \qquad \chi'(x_0)=0. \label{cond_fixed_n}
\end{equation}
Stability conditions at the stationary point $(x_0,0)$ read
\begin{align}
\frac{\partial y'_\tau}{\partial x} \bigg \vert _{(x,y)=(x_0,0)} &= -\frac{ \chi ''(x_0)}{1+ \chi (x_0)}<0, \label{stability_cond_n1}\\
\frac{\partial y'_\tau}{\partial y} \bigg \vert _{(x,y)=(x_0,0)} &= - \sqrt{3 \chi (x_0)}<0. \label{stability_cond_n2}
\end{align}
Since the potential $\chi(x)$ is non-negative, from Eq. \Ref{stability_cond_n1} we find that the necessary condition providing stability is $\chi''(x_0)>0$.


\subsubsection*{Nullclines and attractors} 
In general, a nullcline $y'_\tau=0$ is written as follows:
\begin{equation}
\textstyle
-  \sqrt{3} y \left [ 1-y^2+ \chi (x) \right ] \sqrt{[{\textstyle\frac{1}{2}}y^2+\chi (x)]\left(1-{\textstyle\frac{3}{2}}y^2 \right)}
-  \left (1- \frac{3}{2}y^2  \right ) \left (1- \frac{1}{2}y^2  \right ) \chi ' (x) =0.
\label{nknull}
\end{equation}
There is a special solution of Eq.~(\ref{nknull}):
\begin{equation}
\textstyle y = \pm \sqrt{\frac{2}{3}}.
\label{nkas1}
\end{equation}
The other solutions of Eq.~(\ref{nknull}) depend on the functional form of $\chi (x)$.
In case $\chi(x)\to\pm\chi_\pm<\infty$ and $\chi'(x)\to 0$ in the limit $x\to\pm \infty$, then Eq.~(\ref{nknull}) has the solutions $y=0$ and $y = \pm \sqrt{2/3}$.
The stable line of them for the trajectories is only $y=0$, so $y=0$ is only the attractor in the limit $x \rightarrow \pm \infty$
if $\chi (\pm \infty)$ converges. The stability of the lines are evaluated by the sign of $dy/d \tau$ in the region
$0 < \vert y \vert < \sqrt{2/3}$.
On the other hand, the nullclines in the limit $\vert \chi (x) \vert \gg 1$ fulfill Eq.~(\ref{nkas1}) or
\begin{equation}
y \chi ^{\textstyle\frac{3}{2}}(x) \simeq
-\sqrt{3(1- \frac{3}{2}y^2)} \left ( 1- \frac{1}{2}y^2 \right ) \chi ' (x).
\label{nkas2}
\end{equation}
If $y \neq \pm \sqrt{2/3}$, the behavior of the nullclines (\ref{nkas2}) are determined by
the large/small relation between $\chi ^{\textstyle\frac{3}{2}}$ and $\vert \chi ' \vert$.
If $\chi ^{\textstyle\frac{3}{2}} \gg \vert \chi ' \vert$ in the limit $x \rightarrow \pm \infty$, Eq.~(\ref{nkas2}) yields
\begin{equation}
y = -\frac{\chi ' (x)}{\sqrt{3}  \chi ^{\textstyle\frac{3}{2}}(x)}.
\label{nkas3}
\end{equation}
Equation (\ref{nkas3}) is a stable curve for the trajectories,
because $dy/ d \tau$ is negative for positive $y$ and $dy/ d \tau$ is positive for negative $y$.
However, Eq.~(\ref{nkas3}) is not an asymptote of the trajectories, because it is seen that the nullcline \Ref{nkas3} only
exist in the second quadrant and in the fourth quadrant
by taking into account $\chi ' (+\infty) >0$ ($\chi ' (-\infty) <0$) caused from $\chi (+ \infty) \rightarrow + \infty$
($\chi (- \infty) \rightarrow + \infty$).
The nullclines (\ref{nkas1}) are unstable in this case, therefore, there is no attractors in the limit $x \rightarrow \pm \infty$
if $0 \neq \vert \chi ' \vert \ll \chi ^{\textstyle\frac{3}{2}}$.
Else if $\chi ^{\textstyle\frac{3}{2}} \ll \vert \chi ' \vert$ holds,
the asymptotes of the nullcline (\ref{nkas2}) is given by
\begin{equation}
y = - \sqrt{\frac{2}{3}} + \frac{9}{2 \sqrt{6}} \frac{\chi ^3 (x)}{\chi ^{\prime 2} (x)}.
\label{23}
\end{equation}
The curve (\ref{23}) is stable in $y$ direction for the trajectories, moreover, the trajectories on the curve go toward $x \rightarrow - \infty$
if $\chi (- \infty) \rightarrow  +\infty$, because the nullcline (\ref{23}) not only exist in the fourth quadrant but also exist in the third quadrant.
Therefore, Eq.~(\ref{23}) is the asymptote of the trajectories if $\chi ^{\textstyle\frac{3}{2}} \ll \vert \chi ' \vert$.
If $ \chi '  \propto \chi ^{\textstyle\frac{3}{2}}$, there is one solution of $y$ in Eq.~(\ref{nkas2}).
The signature of the solution is negative, therefore, it also becomes the asymptote in the third quadrant.
All kinds of the attractors in the case $\eta <0$ is summarized in Table~\ref{table3}.

\subsection*{The case $\eta = 0$}
{ The case of $\eta =0$ is equivalent to the quintessence model which is well investigated. For example, in Ref. \cite{hep-th:0202017} cosmological evolution has been investigated in models with an arbitrary effective potential $V(\phi)$, including the situation when $V(\phi)$ may become negative for some values of the field $\phi$. Note that in case $\eta = 0$ our results are coincided with those obtained in \cite{hep-th:0202017} for positively defined potentials $V(\phi) > 0$. Nevertheless, for the sake of completeness we shortly review the cosmological dynamics in this case.}
By defining the dimensionless variables,
\beq
x = \sqrt{\pi }\phi,\quad
y = \sqrt{\pi}  \, \dot \phi, \quad
\tau = t, \quad
\chi(x) = \pi V(x/\sqrt{\pi}),
\eeq
we can rewrite Eq. \Ref{phi2gen} as the following autonomous dynamical system:
\begin{align}
\frac{dx}{d \tau} &=
y,
\label{A1v} \\
\frac{dy}{d \tau} &
=-2y\sqrt{3(y^2+2\chi(x))}-\chi'(x).
\label{A2v}
\end{align}
Stationary points $(x_0,y_0)$ of the system (\ref{A1v})-(\ref{A2v}) are determined as follows:
\begin{equation}
y_0=0, \qquad \chi'(x_0)=0. \label{cond_fixed_v}
\end{equation}
Stability conditions for the stationary point $(x_0,0)$ are
\begin{align}
\frac{\partial y'_\tau}{\partial x} \bigg \vert _{(x,y)=(x_0,0)} &= -\chi ''(x_0) <0, \label{stability_cond_v1} \\
\frac{\partial y'_\tau}{\partial y} \bigg \vert _{(x,y)=(x_0,0)} &= -2 \sqrt{6 \chi (x_0)}<0. \label{stability_cond_v2}
\end{align}
These conditions explicitly show that a local minimum of the potential, where $\chi''(x_0)>0$, is stable, while a local maximum, where $\chi''(x_0)<0$, is unstable.

Equation (\ref{A2v}) shows that the nullcline should satisfy $y \chi ' <0$ or $y =\chi ' =0$.
Then the condition $\chi >0$ impose the nullcline to be in the second quadrant or fourth quadrant in the
limit $x \rightarrow \pm \infty$.
Therefore, the nullclines cannot be the attractor in the region $x \rightarrow \pm \infty$ except for the case they are the
stationary points.
While, the nullcline $dy/d\tau =0$ satisfy the following condition if $\chi (x)>0$:
\begin{equation}
\textstyle
y^2 = - \chi (x) + \sqrt{\chi ^2 (x) + \frac{1}{12}\chi'^2(x)}.
\label{k0as}
\end{equation}
Equation (\ref{k0as}) shows that there are one negative solution and one positive solution with respect to $y$.
Therefore, there is only one nullcline in the limit $x\rightarrow +\infty$ ($x\rightarrow -\infty$).
While, Eq.~\Ref{A2v} says that sgn$(dy/d \tau )$ = $-$sgn$(y)$ is satisfied in the limit $y^2 \gg \chi , \vert \chi ' \vert$,
so the nullcline is the only candidate for the attractor in the region $x \rightarrow \pm \infty$ or $y \rightarrow \pm \infty$
except for the stationary points.
Because of the reasons above, there are no attractors in the region $x \rightarrow \pm \infty$ or $y \rightarrow \pm \infty$
except for the stationary points

\section{Slow-roll regime}
As is well known, the usual slow-roll conditions for the minimally coupled scalar field read
$\frac12 \dot\phi^2\ll V$
and
$\ddot\phi \ll 3H\dot\phi.$
Here, the first condition means that the kinetic energy is much less than the potential one, while the second condition says that the `viscosity' given by the term $3H\dot\phi$ is very high.
In the slow-roll approximation, the field equations are reduced to
$3H^2\simeq 8\pi V(\phi)$ and $\dot\phi\simeq-\frac{V_\phi}{3H}$,
and, as the consequence, one additionally has
$-{\dot H}/{H^2}\ll 1$. The last condition provides an exponential (de Sitter) expansion of the Universe.

In the theory \Ref{action} with the nonminimal kinetic coupling, the field equations have the modified form \Ref{genfieldeq}, and hence the slow-roll conditions should be also modified. Let us consider the equation \Ref{00cmpt}. Now, the condition $\frac12\dot\phi^2\ll V(\phi)$ has to be changed as follows
\beq\label{msr1}
\dot\phi^2|1-9\eta H^2|\ll 2V(\phi),
\eeq
so that from Eq. \Ref{00cmpt} we obtain
\beq\label{slowrollrel1}
3H^2\simeq 8\pi V.
\eeq

For the modified equation \Ref{eqmocosm} instead of $\ddot\phi \ll 3H\dot\phi$ we should impose the following condition:
\beq\label{modviscosity}
\ddot\phi\left|(1-3\eta H^2)\right|\ll 3H\dot{\phi}\left|[1-\eta(2\dot H+3 H^2)]\right|.
\eeq
Imposing also the usual requirement onto the rate of the Hubble parameter:
\beq\label{msr2}
\frac{|\dot H|}{H^2}\ll 1,
\eeq
we find 
\beq\label{modviscos}
\ddot\phi\left|(1-3\eta H^2)\right|\ll 3H\dot{\phi}\left|(1-3\eta H^2)\right|.
\eeq
In the case $3\eta H^2\not\equiv 1$ the last inequality reduces to the standard slow-roll condition
\beq\label{msr3}
\ddot\phi \ll 3H\dot{\phi}. 
\eeq
Now, the field equation \Ref{eqmocosm} can be written as 
\beq\label{dotphi}
3H\dot{\phi} \simeq -\frac{V_\phi}{1-3\eta H^2} \simeq -\frac{V_\phi}{1-8\pi\eta V}.
\eeq
The modified equation \Ref{11cmpt} together with the conditions \Ref{msr1}, \Ref{msr3}, and \Ref{msr4} gives us an additional condition
\beq\label{msr4}
6|\eta|H^2\dot{\phi^2}\ll V.
\eeq
Finally, we have got three modified slow-roll conditions: \Ref{msr1}, \Ref{msr2}, and \Ref{msr4}.

With the help of the obtained slow-roll conditions, we can recast the usual slow-roll parameters $\epsilon _H$ and $\eta _H$ in terms of requirements on the derivatives
of the scalar potential:
\begin{align}
\epsilon _ H = -\frac{\dot H}{H^2} & \simeq \frac{V_\phi ^2}{16 \pi V^2 (1-8 \pi \eta V)},
\label{epsilonH}
\\
\eta _H = - \frac{1}{2} \frac{\ddot H}{H \dot H} & \simeq - \frac{\ddot \phi}{H \dot \phi} - \frac{3 \eta H^2}{1-3 \eta H^2} \epsilon _H \nonumber \\
& \simeq \frac{V_{\phi \phi}}{8 \pi V (1- 8 \pi \eta V)} - \frac{V_\phi ^2 (1-16 \pi \eta V)}{16 \pi V^2 (1- 8 \pi \eta V)^2}.
\label{etaH}
\end{align}
Note that the relations (\ref{epsilonH}) and (\ref{etaH}) coincide with those obtained in \cite{Yang:2015pga}.

\begin{figure}
		\begin{center}
			\includegraphics[clip, width=0.6\columnwidth]{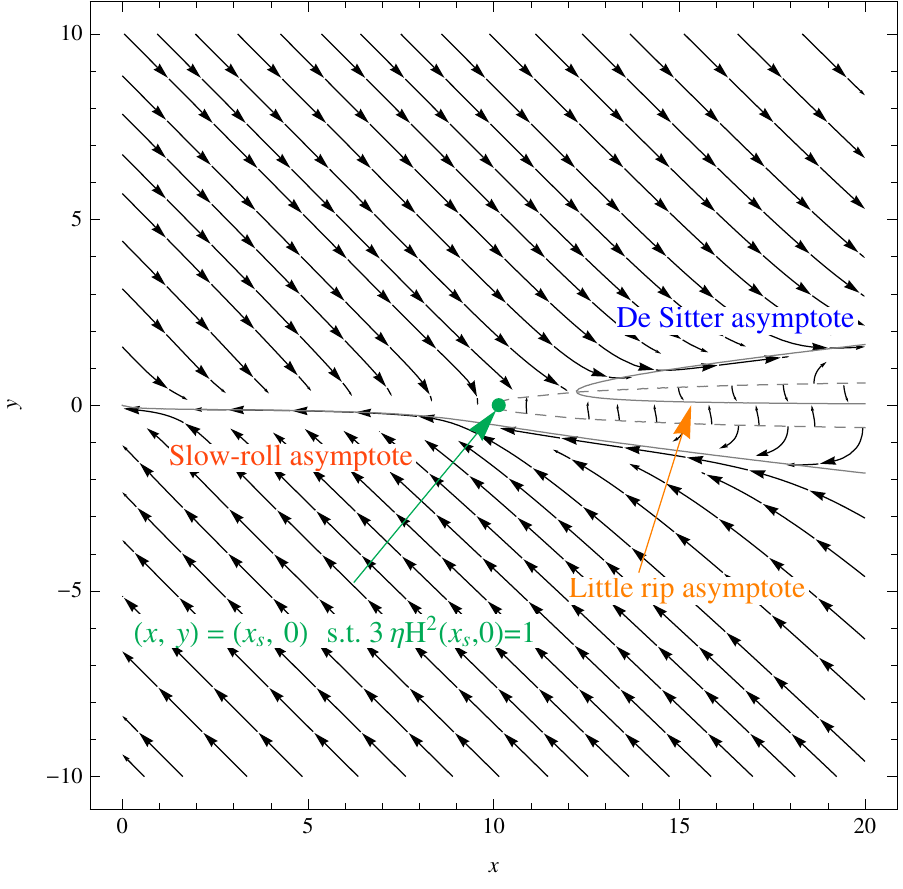}
		\end{center}
	\caption{The phase diagram for $\chi(x)=\frac12\mu^2 x^2$ with $\mu^2=0.02$.
		Solid gray curves and a dashed gray curve represent nullclines $y'_\tau=0$ and the critical curve $y'_\tau = \pm \infty$, respectively. 
		There are three kinds of asymptotes; de Sitter asymptote, little rip asymptote, and slow-roll asymptote. They are 
		stable nullclines, and are separated each other by the critical curve $y'_\tau = \pm \infty$. 
	}
	\label{sr}
\end{figure}

Let us consider the standard square-law potential $V=\frac12 m^2\phi^2$ to illustrate the use of the slow-roll approximation in the theory of gravity with the nonminimal kinetic coupling. From Eqs. \Ref{epsilonH} and \Ref{etaH} we find
\beq\label{srparam}
\epsilon_H \simeq \frac{1}{4\pi\phi^2(1-4\pi\eta m^2 \phi^2)}, \quad
\eta_H \simeq \frac{\eta m^2}{(1-4\pi\eta m^2 \phi^2)^2}.
\eeq
Let $\eta$ be positive, i.e. $\eta>0$. Then, it is obvious that the slow-roll conditions $|\epsilon_H|\ll 1$ and $|\eta_H|\ll 1$ are violated if $4\pi\eta m^2 \phi^2=1$.
The violation of the slow-roll conditions is, in fact, caused by the existence of the singular curves $y'_\tau = \pm \infty$, which appear when 
the conditions $y^2 =0$ and $3 \eta H^2 =1$ are satisfied (see Fig. \ref{sr}). 
In other words, we can say that the singular curves $y'_\tau = \pm \infty$ divide the slow-roll region into some pieces, where either 
$3 \eta  H^2 \ll 1$ or $3 \eta H^2 \gg 1$ is satisfied. 
In the following, we will explore these cases. 

\textbf{The case $4\pi\eta m^2 \phi^2\ll 1$.} In this case one gets $\epsilon_H\simeq \frac{1}{4\pi\phi^2}$ and $\eta_H\simeq \eta m^2$. Introducing the mass parameter $m_\eta=\eta^{-1/2}$, which corresponds to the dimensional coupling parameter $\eta$, we can represent the slow-roll condition $|\eta_H|\ll 1$ as
\beq
\mu^2\equiv\left(\frac{m}{m_\eta}\right)^2\ll 1,
\eeq
where we have introduced the dimensionless parameter $\mu=m/m_\eta$, which characterizes a ratio of scalar and `non minimal coupling' masses.
The slow-roll condition $|\epsilon_H|\ll 1$ together with the demand $4\pi\eta m^2 \phi^2\ll 1$ gives the interval of values of the scalar field for which the slow-roll regime is realized: 
\beq\label{valuephi}
1 \ll 4\pi\phi^2 \ll \frac{1}{\mu^2}.
\eeq
Note that in the case $4\pi\eta m^2 \phi^2\ll 1$ the relation \Ref{dotphi} reduces to
$3H \dot{\phi} \simeq -m^2\phi.$
Thus, the derivative of the scalar field is negative, i.e. $\dot\phi<0$, and hence a value of scalar field $\phi$ is decreasing during the considered slow-roll regime. It means that in some time the condition \Ref{valuephi} will be violated. This moment of time corresponds to the end of the slow-roll regime and transition to the oscillating phase near the minimum of the scalar potential $V(\phi)=\frac12 m^2\phi^2$.

For purposes of numerical analysis it is convenient to use the dimensionless variables \Ref{dlp}, in which the scalar potential takes the form $\chi(x)=\frac12\mu^2 x^2$. In the next section we discuss some specific forms of the scalar potential and, in particular, we consider two examples of the square-law potential with $\mu^2=2$ and $\mu^2=0.02$ (see Figs. \ref{pl} and \ref{sr2}). The numerical analysis clearly confirms that the standard slow-roll phase is absent if $\mu^2=2$ and present if $\mu^2=0.02$.

\textbf{The case $4\pi\eta m^2 \phi^2\gg 1$.} In this case the slow-roll parameters \Ref{srparam} take the following form:
\beq
\epsilon_H\simeq -\frac{1}{(4\pi\mu\phi^2)^2}, \quad
\eta_H\simeq \frac{1}{(4\pi\mu\phi^2)^2}.
\eeq
The conditions $|\epsilon_H|\ll 1$ and $|\eta_H|\ll 1$ are obviously fulfilled for large enough values of $\phi$. In the case $4\pi\eta m^2 \phi^2\gg 1$ the relation \Ref{dotphi} reduces to 
\beq
3H \dot{\phi} \simeq \frac{1}{4\pi\eta\phi}. 
\eeq
It means that for large $\phi$ the derivative $\dot\phi$ is positive, and hence a value of scalar field is continuing to be monotonically increasing and the conditions $|\epsilon_H|\ll 1$ and $|\eta_H|\ll 1$ become only stronger during the slow-roll phase. Therefore, one may expect that such the type of a slow-roll behavior would represent a late time evolution of the Universe with the Hubble parameter which satisfy the conditions $|\dot H/H^2|\ll 1$ and $|\ddot H/2 H \dot H|\ll 1$. Particularly, in the next section it will be shown that for the square-law potential two such scenarios are realized: quasi-de Sitter expansion and a Little Rip. 

\section{Specific examples \label{sec4}}
In this section we apply the general consideration given above for some specific potentials. Namely, we will consider three examples: power-law, Higgs-like, and exponential potentials. It is worth noticing that cosmological dynamics in the theory of gravity with the scalar field possessing a nonminimal kinetic coupling to gravity and the power-law potential has been investigated in Ref. \cite{SkuSusTop:2013}.

\subsection{Power-law potential \label{sec4a}}
Let us consider the power-law potential
\begin{equation}\label{power-law-potential}
V(\phi)=M^4\left(\frac{\phi}{\phi_0}\right)^\sigma,
\end{equation}
where $M$ and $\phi _0$ are positive constants with dimension of mass,
and $\sigma$ is a real number.
Note that cosmological dynamics in the theory of gravity with the scalar field possessing a nonminimal kinetic coupling to gravity and the power-law potential has been investigated in Ref. \cite{SkuSusTop:2013}.

In the dimensionless notation \Ref{dlp} or \Ref{dln}, we obtain
\beq
\chi(x)=\chi_0\, x^\sigma,
\label{ple}
\eeq
where $\chi_0=\frac{8\pi|\eta|M^4}{(8\pi\phi_0^2)^{\sigma/2}}$.
If $\sigma>0$, then the potential \Ref{ple} has the only local minimum at $x=0$, so that $\chi(0)=0$. The point $(x,y)=(0,0)$ on the phase plane is a stationary one if $\chi'(0)=0$, and hence $\sigma>1$. For $\sigma>1$ we have $\chi''(0)>0$, and therefore the stationary point $(0,0)$ is stable regardless of $\eta$.

\begin{figure}
\begin{minipage}[t]{0.5\columnwidth}
\begin{center}
\includegraphics[clip, width=0.97\columnwidth]{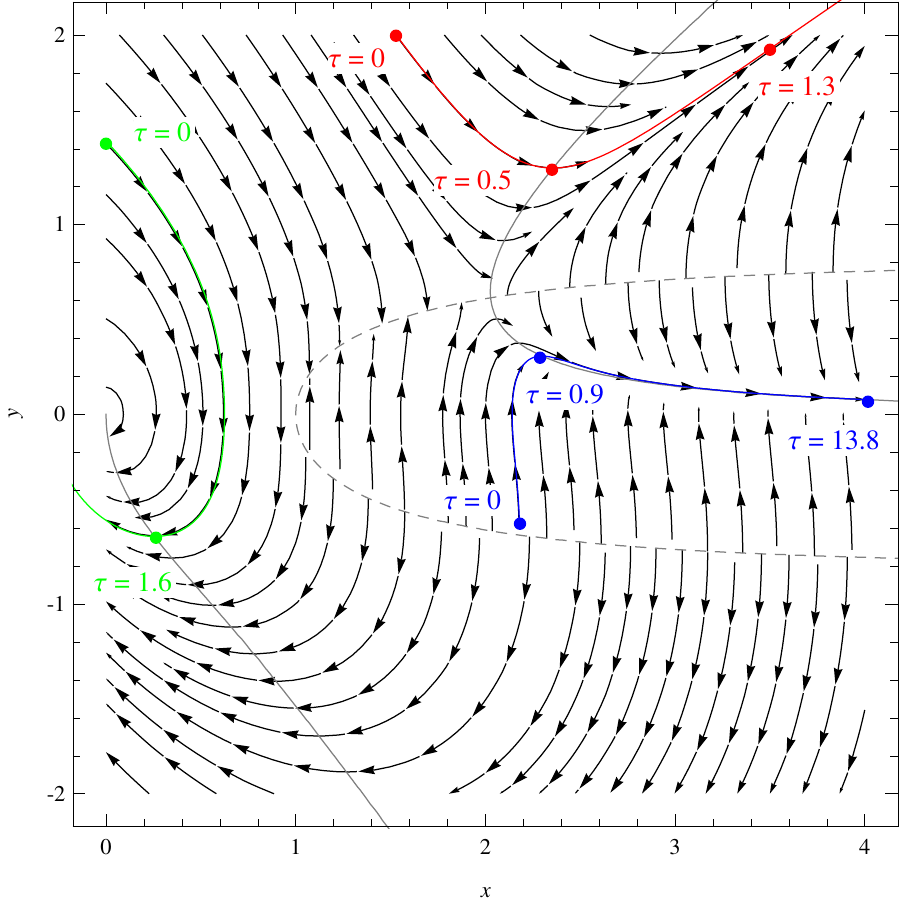}
\end{center}
\end{minipage}%
\begin{minipage}[t]{0.5\columnwidth}
\begin{center}
\includegraphics[clip, width=0.97\columnwidth]{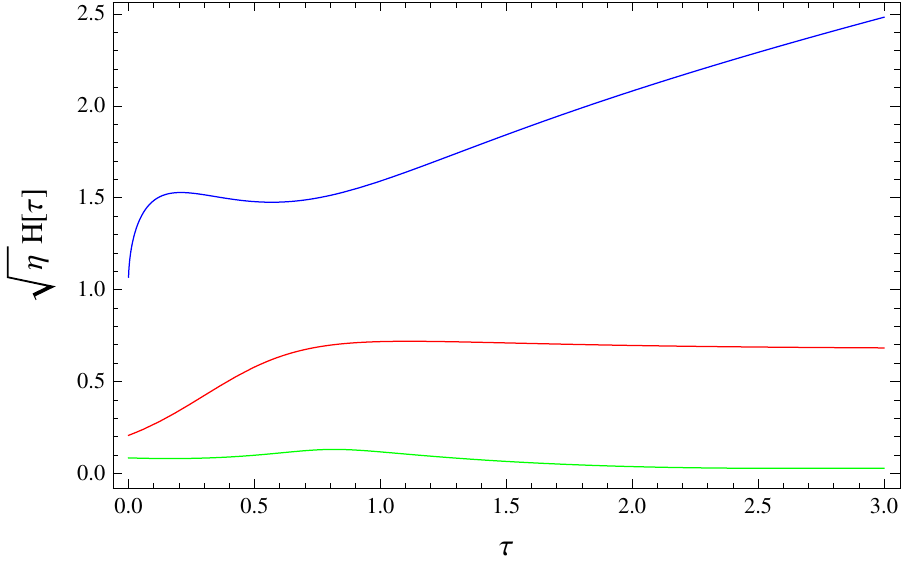}
\end{center}
\end{minipage}
\caption{The phase diagram for $\chi(x)=\frac12\mu^2 x^2$ with $\mu^2=2$.
Solid gray curves represent nullclines $y'_\tau=0$, and the dashed gray curve represents a critical curve $y'_\tau=\pm \infty$.
Examples of the trajectories which approach to attractors are shown as colored curves in the left panel,
and the corresponding time evolution of the Hubble parameter $H(\tau)$ is given in the right panel.}
\label{pl}
\end{figure}

\begin{figure}
	\begin{minipage}[t]{0.5\columnwidth}
		\begin{center}
			\includegraphics[clip, width=0.97\columnwidth]{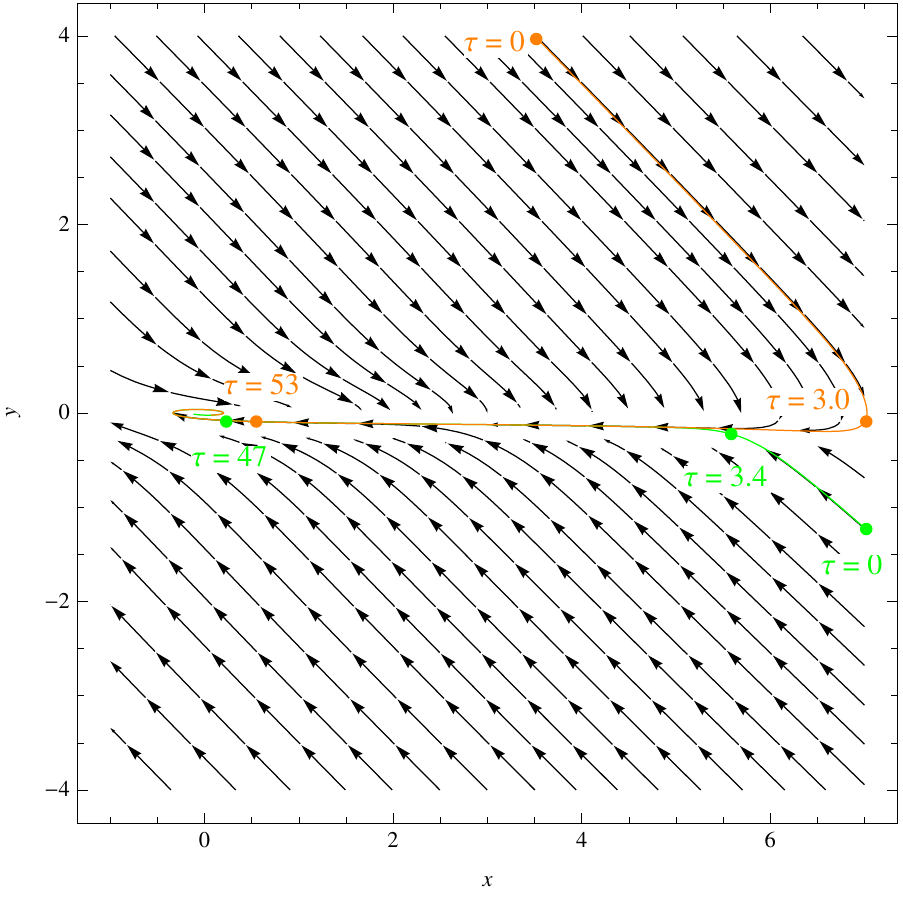}
		\end{center}
	\end{minipage}%
	\begin{minipage}[t]{0.5\columnwidth}
		\begin{center}
			\includegraphics[clip, width=0.97\columnwidth]{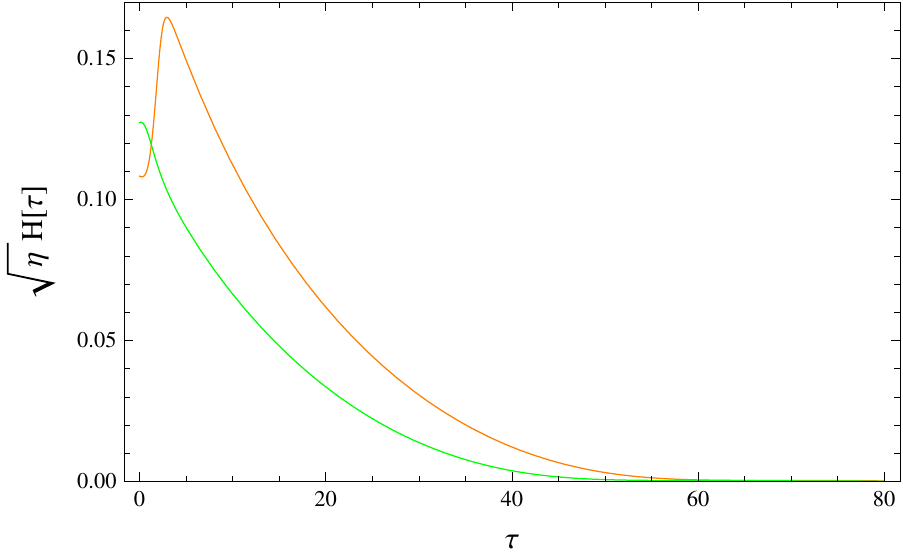}
		\end{center}
	\end{minipage}
	\caption{Slow-rolling trajectories and corresponding time evolutions of the Hubble rate parameter $H(\tau)$ for $\chi(x)=\frac12\mu^2 x^2$ with $\mu^2=0.02$. 
	The trajectories on slow-rolling regime evolve slowly compared to green curve or red curve in Fig. \ref{pl}.
	}
	\label{sr2}
\end{figure}

An asymptotical behavior of $\chi(x)$ depends on $\sigma$. Namely, in the limit $x\to\infty$ one has (i) $\chi$ is finite if $\sigma\leq 0$; (ii) $\chi\to\infty$ and $|\chi'|\ll\chi^{1/2}$ if $0<\sigma<2$; (iii) $\chi\to\infty$ and $|\chi'|\sim \chi^{1/2}$ if $\sigma=2$; (iv) $\chi\to\infty$ and $\chi^{1/2} \ll |\chi'| \ll \chi^{3/2}$ if $\sigma>2$.

Now, basing on the above analysis (see the table \ref{table2}), we can conclude that in the limit $x\to\infty$ there exists an attractor $y=0$ if $\sigma<0$; 
the corresponding solution is  $\phi(t) = \phi_\infty=const$.
If $\sigma=0$, 
there exists the asymptote $y=y_{\infty}=\sqrt{8\pi\eta M^4-1}$
provided $8\pi\eta M^4>1$; the corresponding solution is $\phi(t)=(y_\infty/\sqrt{8\pi\eta})\,t+const$.
The second asymptote $y=-\chi'/[\sqrt{3\chi}(1-\chi)]=0$ is neither stable not unstable because $\chi'\equiv 0$.
In the case $0<\sigma<2$ 
there are two asymptotes. The first one is $y=\chi'/\sqrt{3\chi^3}$
with the asymptotical solution
\begin{equation}
\phi(t)=\left[\frac{\sigma(\sigma +4)}{16\pi\eta}\sqrt{\frac{\phi_0^\sigma}{24\pi M^4}}\,t +const  \right]^{\frac{2}{\sigma+4}}.
\label{420}
\end{equation}
The corresponding asymptote for the Hubble parameter in the limit $\phi_0 t\to\infty$  is $H(t) \propto t ^{\sigma/(4+\sigma)}$.
Such the behavior of $H(t)$ means the so-called Little Rip \cite{Frampton:2011sp}, which represents an expansion faster than de Sitter one.
The second asymptote is given as $y=\sqrt{\chi}$ with the asymptotical solution
\begin{equation}
\phi (t) = \left[\frac{(2-\sigma)M^2}{2\phi_0^{\sigma/2}}\, t
+const\right]^{\frac{2}{2-\sigma}}.
\label{430}
\end{equation}
The relation \Ref{430} shows that $\phi$ is an increasing function of time.
At the same time, the corresponding asymptote for the Hubble parameter is $H=1/\sqrt{3 \eta} = const$.
In the case $\sigma =2$ 
one also has two asymptotes in the limit $x\to\infty$: $y=\chi'/\sqrt{3\chi^3}$ and $y=\beta\sqrt{\chi}$, where $0<\beta <1$. Qualitatively, the asymptotical behavior of $H(t)$ is the same as that in the case $0<\sigma<2$.
In the case $\sigma>2$ there exist two asymptotes: $y=\chi'/\sqrt{3\chi^3}$ and $y=\beta\sqrt{\chi^3}/\chi'$. The parameter $\beta$ is determined as $\beta=\frac{4\sqrt{2}\sigma}{3\sigma+2}$, and hence its value is constrained within the range $\sqrt{2}<\beta<4\sqrt{2}/3$.
A solution on the asymptote $y=\beta\sqrt{\chi^3}/\chi'$ is given as follows:
\begin{equation}
\phi(t)=\phi _0 \left[const -\frac{\beta\sigma}{2}\sqrt{8\pi}M^2 t \right]^{-{2}/{\sigma}}.
\label{460}
\end{equation}
It is worth noticing that at a some moment of time $t_{*}$ the scalar field $\phi(t)$, given by Eq.~\Ref{460}, becomes singular. The corresponding Hubble parameter, given as
$H=\sqrt{2V_\phi/(9\sigma(\eta V)^{1/2})}$, also goes to infinity at the moment of time $t_*$. Such the singularity is known as the Big Rip \cite{Big}.

Summarizing, we can conclude that the model with the power-law potential \Ref{power-law-potential} provides three types of the accelerating cosmological expansion in the limit $\phi\to\infty$: (i) the de Sitter expansion; (ii)
the Little Rip; and (iii) the Big Rip. Also, there exists an attractor $\phi\to 0$ and $\phi'\to 0$; the Hubble parameter is approaching to this attractor with damping oscillations. In the figure \ref{pl} we give a numerical illustration of various cosmological scenarios in the case of the quadratic potential with $\sigma=2$.

\subsection{Higgs-like potential}
Now, let us consider a Higgs-like potential
\begin{equation}
V(\phi) = \frac{\lambda}{4} (\phi ^2 - \phi _0 ^2)^2,
\label{470}
\end{equation}
where $\lambda$ is a positive dimensionless constant and $\phi _0$ is a positive constant with dimension of mass. The detailed consideration of cosmological dynamics  with the nonminimal kinetic coupling and the Higgs-like potential has been given in Ref. \cite{Matsumoto:2015hua}.

\begin{figure}
\begin{minipage}[t]{0.5\columnwidth}
\begin{center}
\includegraphics[clip, width=0.97\columnwidth]{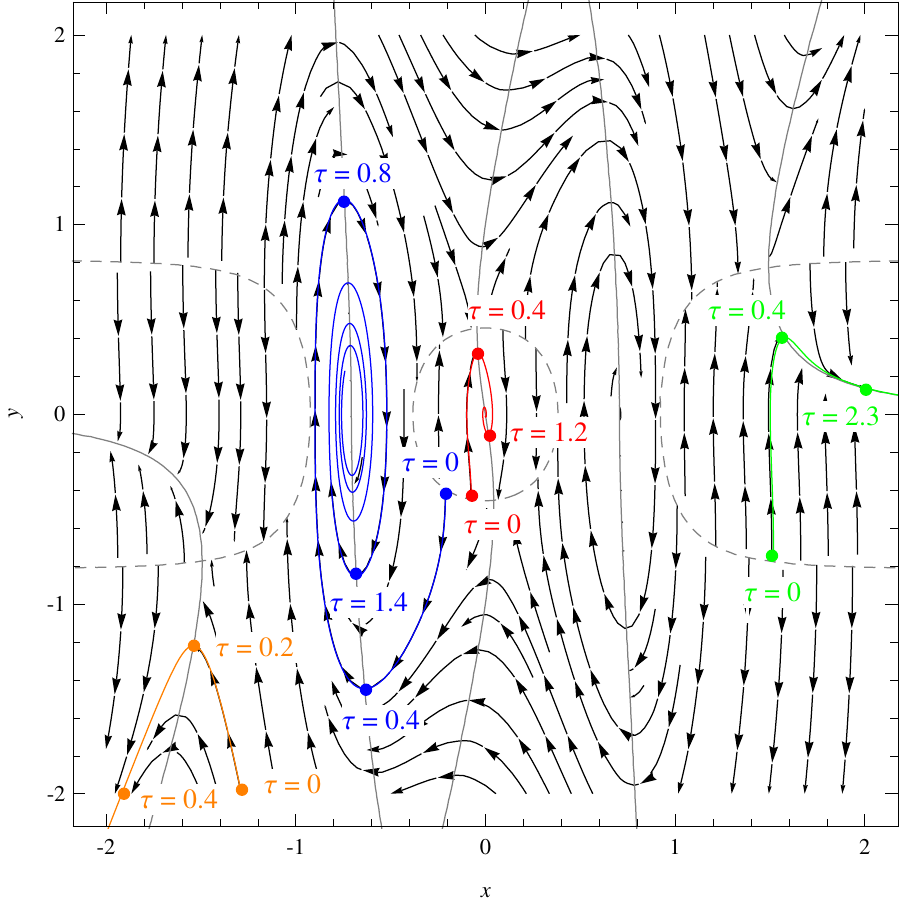}
\end{center}
\end{minipage}%
\begin{minipage}[t]{0.5\columnwidth}
\begin{center}
\includegraphics[clip, width=0.97\columnwidth]{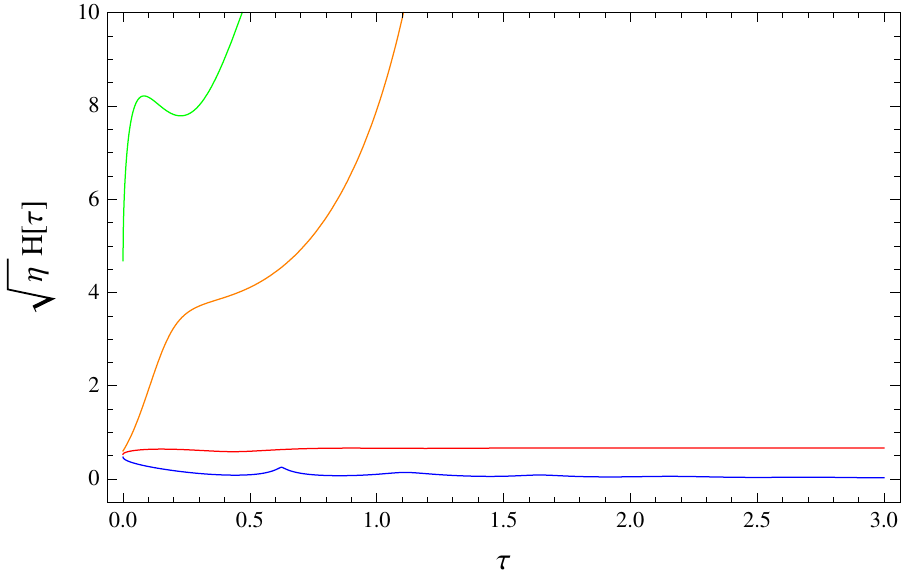}
\end{center}
\end{minipage}
\caption{A phase diagram in the case $\chi (x) = 8(x^2-0.5)^2$.
Solid gray curves and a dashed gray curve represent $dy/d \tau =0$ and $dy/d \tau = \pm \infty$, respectively.
Examples of the trajectories which go to the attractors are shown as colored curves in the left panel,
and the corresponding time evolutions of the Hubble parameter $H(t)$ are expressed in the right panel. }
\label{higgs}
\end{figure}

The stationary points located in the restricted region are given by $(x,y)=(0,0)$,
$(\pm x_0,0)$, where $x_0=\sqrt{8\pi}\,\phi_0$.
Values of the potential at the stationary points are $V(\pm\phi_0)=0$ and
$V(0)=\lambda\phi_0^4/4$, i.e. $\chi(0)=2\pi|\eta|\lambda\phi_0^4$.
The stationary points $(x,y)=(\pm x_0,0)$ are stable, because the conditions
$\chi(x_0)=0$ and $\chi''(x_0) >0$ are satisfied.
On the other hand, the stability of the point $(x,y)=(0,0)$ depends on the value of $\chi (0)$ in case $\eta >0$.
If $\chi (0) >1$, then the point $(x,y)=(0,0)$ is stable, otherwise it is unstable.
A phase diagram in the case of $\chi (0)>0$ is shown in Fig.~\ref{higgs}.
Note that the phase diagram is centrally symmetrical.
In other words,
the first (second) quadrant and the third (fourth) quadrant are completely same except for the direction of the arrows are opposite,
because the potential \Ref{470} is an even function.
There are two types of attractors in $x \rightarrow \pm \infty$. One of them causes
the Little Rip expansion of the Universe, and another causes the Big Rip expansion.
The numbers of attractors are seven in the whole phase diagram if $\chi (0)>1$.
However, the trajectories on the phase diagram are not so complicated,
because the curve $dy/d \tau = \pm \infty$ divides the whole phase diagram into four pieces, and there are no trajectories that cross over the curve $dy/d \tau = \pm \infty$.

\subsection{Exponential potential}
In this section we consider the scalar potential $V(\phi)$ in the Liouville, i.e. exponential form
\begin{equation}
V(\phi)= M^4 \mathrm{e}^{-\phi/ \phi _0},
\label{480}
\end{equation}
where $M$ and $\phi _0$ are positive constants with dimension of mass.
Note that the exponential potential has been considered in numerous papers devoted to cosmological models with scalar fields (see, for instance, \cite{exp}). However, a role of the exponential potential in the theory of gravity with nonminimal kinetic coupling is not studied.

Using Eqs. \Ref{dlp} and \Ref{dln}, we can define the dimensionless potential function $\chi(x)$ as follows:
\beq
\chi(x)=8\pi|\eta|M^4 \mathrm{e}^{-x/x_0},
\eeq
where $x=\sqrt{8\pi}\phi$.
The potential $\chi(x)$ has no extrema. The only stationary point is $(x,y)=(\infty, 0)$.
However, since the relations $\chi(\infty)=\chi'(\infty)=\chi''(\infty)=\cdots =0$ are fulfilled, a stability analysis of the stationary point $(\infty, 0)$ is problematic.
Instead, we will use the analysis we have done in the region $ \vert x \vert \gg 1$.
Equation \Ref{480} yields $0< \chi (x) \ll 1$ and $\chi '(x) < 0$ for
$1 \ll \vert x \vert < + \infty$, so it is seen that the point $(x,y)=(+ \infty,0)$ is an attractor.
To investigate a behavior of trajectories in the region $1 \ll x < + \infty$,
we will impose the conditions $\vert y \vert, \chi , \vert \chi ' \vert \ll 1$ on Eq.~(\ref{A2}).
Then we have
\begin{equation}
\frac{dy}{d \tau} \simeq - \sqrt{3} y \sqrt{\frac{1}{2} y^2 + \chi (x)} - \chi ' (x).
\label{490}
\end{equation}
It is seen that $dy/d \tau >0$ is always satisfied if $y<0$ by taking into account $\chi ' (x) <0$.
Therefore, there is a stable nullcline in $y>0$.
This nullcline is the asymptote leading to the attractor $(x,y) = (+ \infty , 0)$.
The nullcline is given by
\begin{equation}
y = \sqrt{- \chi + \sqrt{\chi ^2 + \frac{3}{2}\chi ^{\prime 2}}}.
\label{4100}
\end{equation}
Taking into account that $y=dx/d \tau$ and $V(\phi)= M^4 \mathrm{e}^{-\phi/ \phi _0}$, we can resolve Eq.~(\ref{4100}) with respect to $x (\tau)$ as follows
\begin{equation}
x(\tau ) = 2x_0 \ln \left [ \frac{1}{2 x_0}
\sqrt {\left ( \sqrt{1+ \frac{3}{2x_0^2}}-1 \right )8 \pi  \eta M^4} \tau + const. \right ],
\label{4110}
\end{equation}
where $x_0 = \sqrt{8 \pi } \phi _0$.
At the same time, the Hubble rate function is given as
\begin{equation}
H(t) = \frac{\gamma}{t},
\qquad \gamma ^2 = \frac{2 x_0^2 \left ( 1+ \sqrt{1+ \frac{3}{2 x_0^2}} \right )}{3 \left ( \sqrt{1+ \frac{3}{2 x_0^2}}-1 \right ) }.
\label{4120}
\end{equation}
If $\gamma = 1/2$, the Hubble rate expressed by Eq.~\Ref{4120} is same as that of the radiation dominant era.
\begin{figure}
\begin{minipage}[t]{0.5\columnwidth}
\begin{center}
\includegraphics[clip, width=0.97\columnwidth]{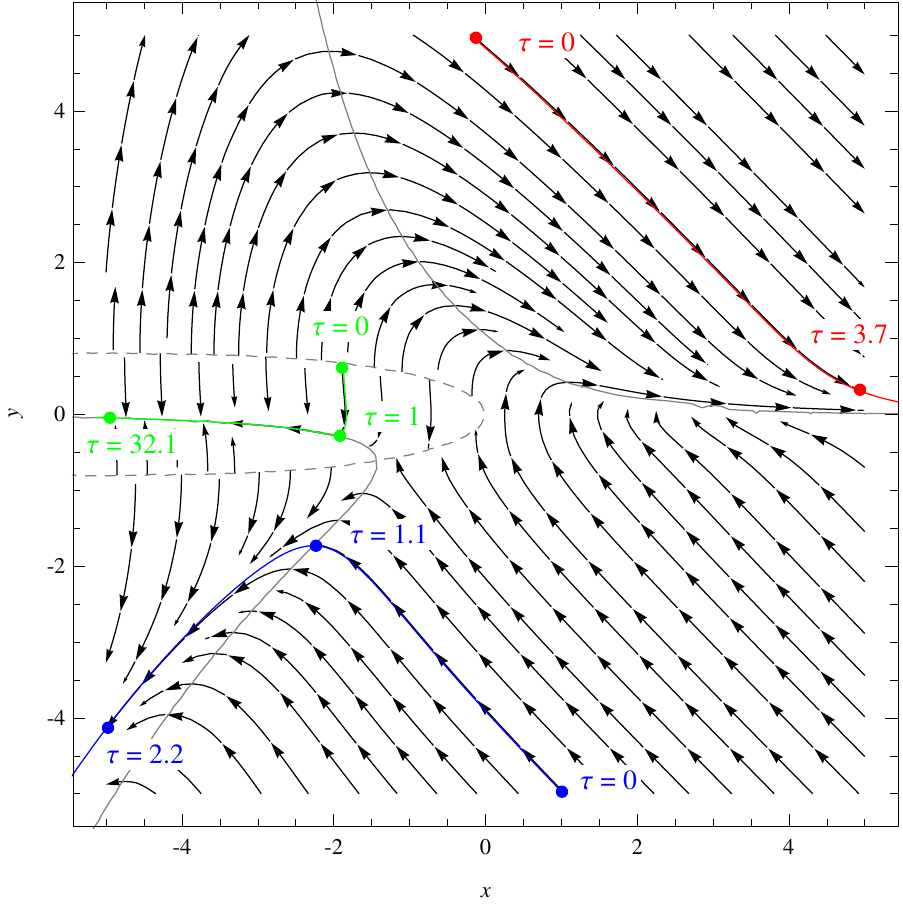}
\end{center}
\end{minipage}%
\begin{minipage}[t]{0.5\columnwidth}
\begin{center}
\includegraphics[clip, width=0.97\columnwidth]{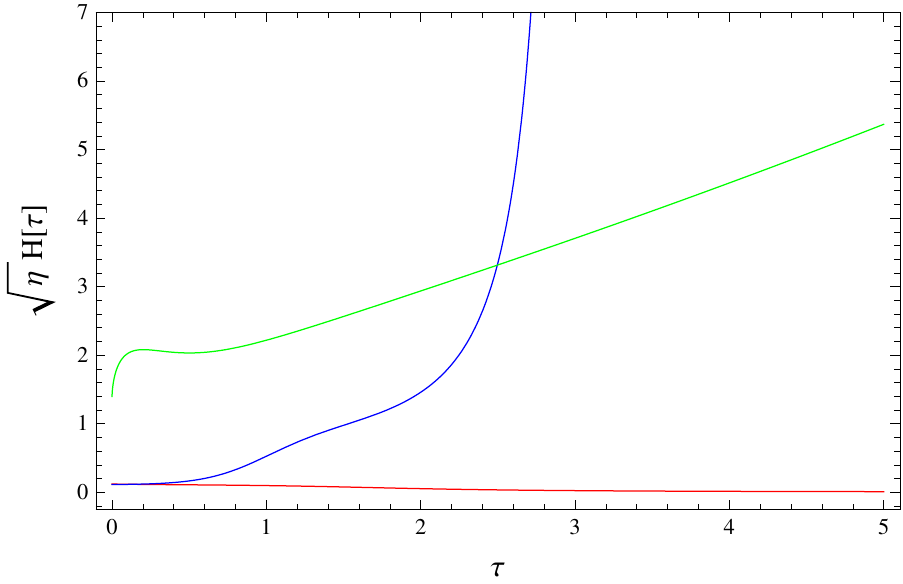}
\end{center}
\end{minipage}
\caption{A phase diagram in the case $\chi (x) = \mathrm{e}^{-x}$.
Solid gray curves and a dashed gray curve represent $dy/d \tau =0$ and $dy/d \tau = \pm \infty$, respectively.
Examples of the trajectories which go to the attractors are shown as colored curves in the left panel,
and the corresponding time evolutions of the Hubble parameter $H(t)$ are expressed in the right panel.}
\label{exp}
\end{figure}

On the other hand, there are the other attractors in the limit $x \rightarrow - \infty$.
For the potential (\ref{480}), $\chi (- \infty) \rightarrow + \infty$ and $\vert \chi ' (- \infty) \vert \sim \chi (- \infty)$
are satisfied.
Therefore, there are two kind of attractors for $x \rightarrow - \infty$ if $\eta > 0$;
$y= \chi ' / \sqrt{3 \chi ^3}$ and $y^2 \propto \sqrt{\chi ^3}/ \chi '$.
Resolving $x' (\tau )= \chi '(x(\tau )) / \sqrt{3 \chi ^3 (x (\tau ))}$ with respect to
$x(\tau )$ gives
\begin{equation}
x(\tau) = -2 x_0 \ln \left [ \frac{1}{4 x_0} \sqrt{\frac{1}{6 \pi  \eta M^4 x_0}} \tau + const. \right ].
\label{4130}
\end{equation}
Therefore, the Hubble rate parameter on the curve $y= \chi ' / \sqrt{3 \chi ^3}$ is
approximately given as
\begin{equation}
H(t) \simeq \frac{t}{6 \eta x_0^{3/2}}.
\label{4140}
\end{equation}
Here, the constant in Eq.~(\ref{4130}) is ignored. The time evolution given by
Eq.~(\ref{4140}) means that the Little Rip expansion is realized on the asymptote $y= \chi ' / \sqrt{3 \chi ^3}$.

The coefficient of the curve $y^2 = \alpha \sqrt{\chi ^3}/ \chi '$ is obtained by equating
$-d [\alpha \sqrt{\chi ^3}/ \chi ']^{1/2}/dx$ with $(dy/d \tau)/ (dx / d \tau)$ on the curve.
Taking the conditions $\vert y \vert \sim \chi ^{1/4} \sim \vert \chi ' \vert ^{1/4} \gg 1$ into account
gives the following value of $\alpha$:
\begin{equation}
\alpha = - \frac{4 \sqrt{2}}{3} > -2 \sqrt{2}.
\label{4150}
\end{equation}
Therefore, the concrete form of the asymptote is
\begin{equation}
y =  -\sqrt{ - \frac{4 \sqrt{2}}{3} \frac{\chi ^{3/2}(x)}{\chi '(x) } }
= - 4 \sqrt{\frac{\sqrt{ \pi  \eta M^4}x_0}{3}} \mathrm {e}^{-\frac{x}{4x_0}}.
\label{4160}
\end{equation}
Then, we can obtain the time evolution of $x$ and $H$ as follows:
\begin{align}
x(\tau) &= 4 x_0 \ln \left [ const. - \sqrt{\frac{\sqrt{\pi  \eta M^4}}{3 x_0}} \tau \right ], \label{4170} \\
H(t) &= \frac{1}{const. -t}. \label{4180}
\end{align}
Eq.~(\ref{4180}) means that the scale factor $a(t)$ and the Hubble rate $H(t)$
go to infinity at some finite time, i.e. the Big Rip occurs.
These characteristic behaviors of the Hubble rate function can be seen in Fig.~\ref{exp}.

\section{Conclusions \label{sec5}}
We have considered a cosmological dynamics of the scalar field with nonminimal kinetic coupling and a potential of general form.
First, we have shown that the field equations
can be represented as an autonomous system by using appropriate dimensionless parameters.
It is worth noticing that the field equations expressed in a dimensionless form contains ultimately only one arbitrary function $\chi (x)= 8 \pi  \vert \eta \vert V(x/\sqrt{8 \pi })$ instead of two arbitrary values, that is the potential $V( \phi )$ and the coupling parameter $\eta$.

Stationary points of the autonomous system are expressed as $(x,y)=(x_0,0)$ provided $\chi '(x_0)=0$ and $\chi (x_0) \neq 1$.
A stability of stationary points is determined by the value of $\chi (x_0)$ and the sign of $\chi ''(x_0)$.
In case $0 \leq \chi (x_0) <1$ a stationary point is stable if $\chi '' (x_0)$ is positive.
In case $\chi(x_0)>1$ it is stable if $\chi''(x_0)$ is negative.
The other attractors of the system are $(x,y)=(\pm \infty , 0)$, $(\pm \infty , L)$,  $(\pm \infty , \pm \infty)$, where $L$ is a finite number.
A behavior of asymptotes which go towards these attractors
has been clarified by investigating critical lines and given in Sec.~\ref{sec3}.
All results are arranged in Tables \ref{table2} and \ref{table3}.
By using the results of Sec.~\ref{sec3}, we have shown that interesting scenarios of  time evolution of the Universe, e.g. de Sitter expansion, the Little Rip and the Big Rip, are realized on the asymptotes given in Sec.~\ref{sec4}.
Existence conditions for the asymptotes depend in general on the form of $\chi (x)$, or, ultimately, $V(\phi)$.

In Sec.~\ref{sec4} some specific forms of the scalar potential $V(\phi)$ have been considered and a behavior of phase trajectories has been evaluated.
As the result, it has been shown that the variety of asymptotes depends  essentially on a power of $\phi$.
In particular, the Hubble rate on the asymptotes becomes the steeper the higher a power of $\phi$.
In the case of exponential potential, it has been shown that there exist generally three attractors which correspond to the decelerated expansion, the Little Rip, and the Big Rip, respectively.

\section*{Acknowledgments}
The work was supported by the Russian Government Program of Competitive Growth of Kazan Federal University.
SVS was supported by the RSF grant 16-12-10401.

\end{document}